\documentclass[journal]{IEEEtran}

\usepackage{cite}
\usepackage{amsmath,amssymb,amsfonts}
\usepackage{algorithmic}
\usepackage{graphicx}
\usepackage{textcomp}
\usepackage{xcolor}
\usepackage{listings}
\usepackage[acronym]{glossaries}
\usepackage{hyperref}
\usepackage{multirow}

\ifCLASSOPTIONcompsoc
\usepackage[caption=false, font=normalsize, labelfont=sf, textfont=sf]{subfig}
\else
\usepackage[caption=false, font=footnotesize]{subfig}
\fi

\def\BibTeX{{\rm B\kern-.05em{\sc i\kern-.025em b}\kern-.08em
    T\kern-.1667em\lower.7ex\hbox{E}\kern-.125emX}}

\newacronym{AD}{AD}{Microsoft Active Directory}
\newacronym{API}{API}{Application Programming Interface}
\newacronym{ASM}{ASM}{Abstract State Machine}
\newacronym{BPM}{BPM}{Business Process Management}
\newacronym{BPMN}{BPMN}{Business Process Model and Notation}
\newacronym{LDAP}{LDAP}{Lightweight Directory Access Protocol}
\newacronym{VPN}{VPN}{Virtual Private Network}
\newacronym{PKI}{PKI}{Public Key Infrastructure}
\newacronym{PASS}{PASS}{Parallel Activity Specification Schema}
\newacronym{AST}{AST}{Abstract Syntax Tree}
\newacronym{OWL}{OWL}{Web Ontology Language}
\newacronym{SBD}{SBD}{Subject Behavior Diagram}
\newacronym{SID}{SID}{Subject Interaction Diagram}
\newacronym{JSON}{JSON}{JavaScript Object Notation}
\newacronym{FIFO}{FIFO}{First In, First Out}
\newacronym{GUI}{GUI}{Graphical User Interface}
\newacronym{ORM}{ORM}{Object-relational Mapping}
\newacronym{XML}{XML}{Extensible Markup Language}
\newacronym{S-BPM}{S-BPM}{Subject-oriented Business Process Management}
\newacronym{IoT}{IoT}{Internet of Things}

\lstdefinestyle{Python} {
	language=Python,
	aboveskip=3mm,
	belowskip=3mm,
	tabsize=3,
	basicstyle={\small\Consolas},
	keywordstyle=\color{blue},
	morekeywords={self},
	float,
	captionpos=b
}

\begin{document}

\title{A Step Towards a Universal Method for Modeling and Implementing Cross-Organizational Business Processes
}

\author
{
\IEEEauthorblockN{Gerhard Zeisler\textsuperscript{\textdagger}},
\and
\IEEEauthorblockN{Tim Tobias Braunauer\textsuperscript{\textdagger}},
\and
\IEEEauthorblockN{Albert Fleischmann},
\and
\IEEEauthorblockN{Robert Singer}
}

\maketitle
\begingroup\renewcommand\thefootnote{\textdagger}
\footnotetext{Equal contribution}
\endgroup

\begin{abstract}
	
The widely recognized Business Process Model and Notation (\acrshort{BPMN}), while prevalent in industry standards for business process modeling, faces limitations in terms of ambiguous execution semantics. This ambiguity can lead to varied interpretations of the execution logic, contingent upon the specific software used for implementation. In contrast, the Process Specification Language (\acrshort{PASS}) is designed to offer formally defined semantics, addressing the interpretational issues inherent in \acrshort{BPMN}. Despite its advantages, \acrshort{PASS} has not achieved the same level of industry adoption as \acrshort{BPMN}.

This feasibility study introduces PASS as an intermediary framework for translating and executing BPMN models. The study involves developing a prototype translator that converts a select subset of BPMN elements into a PASS-compatible format. These translated models can be transformed into source code and executed by our reference implementation of a workflow environment. This approach differs from typical BPMN workflow implementations.

The findings demonstrate that this integrated approach not only enhances compatibility between different modeling and execution tools but also offers a robust methodology for cross-organizational implementation of business processes. This study paves the way for a more unified and precise execution of business process models, potentially revolutionizing the way industries approach process modeling and execution.

\end{abstract}

\begin{IEEEkeywords}
PASS, BPMN, subject-orientation, business process modeling, workflow engine, code generator, cross organizational business processes, agent
\end{IEEEkeywords}


\section{Introduction}
\begin{figure}[htb]
    \includegraphics[width=0.485\textwidth]{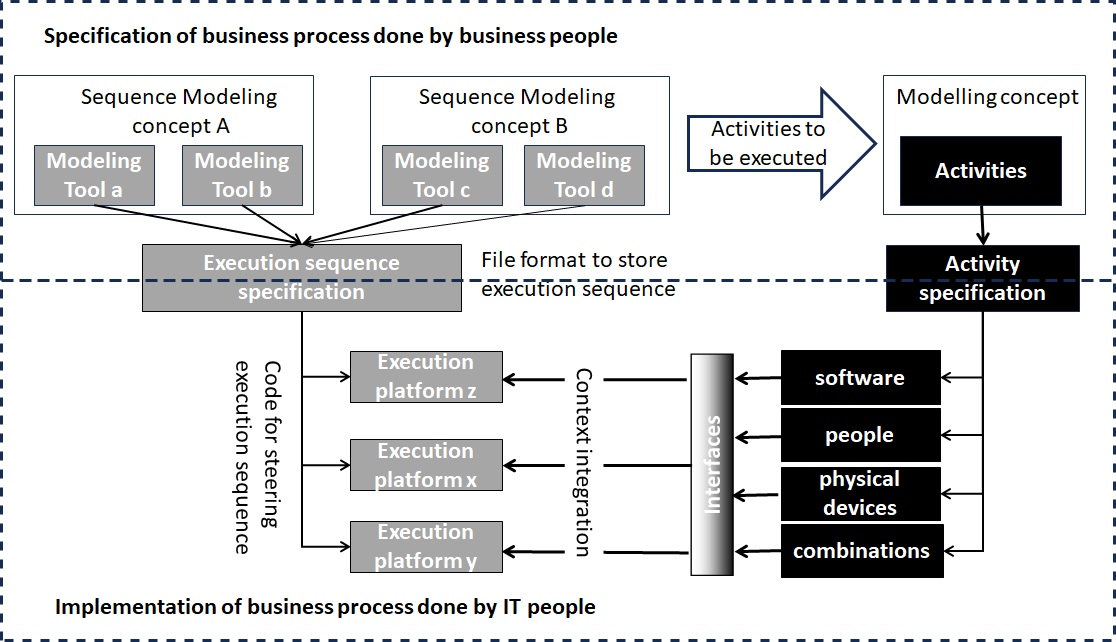}
    \centering
    \caption{Aspects for digitalisation of business processes}
    \label{fig:businessProcessDigitalisation}
\end{figure}

Business processes are highly relevant factors within a modern company and are highly influential on the success of a business. A business process defines the activities and the execution sequence of these activities to produce an intended result. To improve processes and to reduce costs, today many parts of a business process are digitalised. The major aspects of the digitalisation of business processes are shown in \autoref{fig:businessProcessDigitalisation}. There are two major aspects of creating a business process model:
\begin{itemize}
    \item The activities which are executed in a process to get the intended result. This includes the related entities and the operations defined on them. The results of operations executed on objects influence which activity is executed next. 
    \item The sequence in which these activities are executed depending on the results of the operations.
\end{itemize}

Normally both aspects of a process specification are described by business people using appropriate tools. For describing activities, tools that are commonly used in system specifications can be utilised e.g. UML. This aspect is not investigated in this paper. There exists extensive literature about methods and tools to specify functions e.g. \cite{Formal-Methods-Software-Engineering-2023}. For this paper, it is assumed that the meaning of the used activities is described anyhow. In this paper, the focus is on the sequences in which these activities are executed.

To specify the execution sequence of these activities, special description concepts including corresponding tools are developed e.g. EPC, BPMN. The concepts and tools which are used for this work depend on the know-how of the business people.\\
The created models are stored in an appropriate file format and are the input for an execution platform. An execution platform provides the required resources for executing processes. This includes people, physical devices and software systems. \\
Today, processes are defined across organisations. Each involved organisation may have its own execution platform, therefore processes may be executed across various execution platforms.\\
The speciality of business process digitisation is the implementation of the execution sequence the so-called control flow. This aspect is central to implementing business processes. Therefore in this article, this aspect is considered in more detail. \textit{This paper aims to investigate to which extent different specification tools supporting different modelling types can create models which can be understood and executed by various execution platforms.}

In this paper, not all aspects of such a general digitisation environment for business processes are considered. This feasibility study aims to evaluate if it is possible to create business process models with \acrshort{BPMN} based tools and a \acrshort{S-BPM} based tool, and that these models can be executed by two different workflow engines. This enables the utilisation of different modelling tools and different execution systems for the digitisation of a business process. This increases the flexibility and reduces development and maintenance costs.

\acrshort{BPMN} was chosen due to its wide adoption in the industry and the availability of many modelling tools on the market. The storage format for \acrshort{BPMN} models is based on XML but the execution semantic is described in natural language (see \cite{BPMN-OMG-2012}).
S-BPM has been chosen because it is the only business process modelling language for which an \acrshort{OWL} \cite{w3cowlworkinggroupOWLWebOntology2012} based storage format has been defined. Based on that storage \acrshort{OWL} format a formal execution semantic as an \acrlong{ASM} \cite{BoergerABstractStateMachine2003} has been created. 

Previously we have also developed an OWL Ontology  including all elements of the BPMN standard \cite{singer2019ontological}. This ontology has been used to translate BPMN models from BPMN into S-BPM models and vice versa. The developed ontology can also be used to test any BPMN model against the standard. This demonstrates that BPMN models (with conceptual restrictions) can be translated into S-BPM models. For this work, we have decided to directly translate the models based on their origin file structure to have more flexibility during the development cycle, as is explained in \autoref{section:Transformation}.
If a process model is handled by different tools then each tool must understand its syntax and semantics.
\autoref{section:Related-Work} of this paper outlines the related work according to the following questions: Does a storage language for different types of modelling languages already exist? Is there a storage language for which formal execution semantics are defined? Does a workflow engine exist which is able to execute a process logic available in a specific storage format that has been produced by different methods and tools?
In \autoref{section:structure-feasability} an overview of the structure of the feasibility study described in this paper is presented. In \autoref{section:PASS} and \autoref{section:BPMN} \acrshort{BPMN} and \acrshort{S-BPM}/\acrshort{PASS} are shortly described.

Afterwards, in \autoref{section:Transformation}, the \acrshort{BPMN} elements used to model \acrshort{PASS} processes are described. Furthermore, a short description of the translation process is given, including which problems have been encountered and which possibilities the translation process offers.
Afterwards in \autoref{section: workflow-concept}, a concept for a workflow engine based on \acrshort{PASS} models is described. In the subsequent chapter, the concept is implemented as a prototype including a translation from \acrshort{PASS} to Python program code. As a result of the combination with the translation tool, the Python-based workflow engine can execute process models described either in \acrshort{PASS} or \acrshort{BPMN}.
At the end of the paper in \autoref{section:conclusion}, the resulting prototypes as well as the \acrshort{ASM} reference implementation are tested and verified against a specific set of known process patterns in a variety of scenarios involving multiple \acrshort{BPMN} and \acrshort{PASS} modelling tools. 

More details about the concepts described in this article can be found in \cite{BraunauerTimTobias2022BzP:} and \cite{ZeislerGerhard2022Acgf}.
\section{Related work}
\label{section:Related-Work}

In this section, related work to the problem statement of this paper is evaluated. 

\subsection{Does a storage language for different types of modelling languages already exist?}
\label{subsec:storage}
In order to create models with different graphical-based methods two sub-problems have to be solved. The meaning of the symbols in the various languages have to be mapped on each other and based on this mapping the models have to be stored in the same file format. This means the storage format covers the semantics of the various constructs of process modelling languages.

If graphical-based modelling languages are used, which is more or less the standard, the layout of the drawings representing a process model has to be stored.
This means that there must exist a mapping of the graphical symbols to the corresponding data structure of the file format, the positioning of the symbols in the drawings and the alignment of connections between symbols representing a process model.

The authors could not find a storage language that allows storing different types of process models e.g., \acrshort{BPMN} models, flowchart-based models, or Event-driven Process Chain (EPC, EPK) models in the same storage format including the graphical representation. 

There exists an XML-based standard for \acrshort{BPMN} \cite{BPMN-OMG-2012}. This means that different \acrshort{BPMN} modelling tools store models in the same standard format as far as they do not contain vendor-specific modelling elements. Models can be exported from one tool and imported into another tool. But only if the tools follow strictly the standard.
This is possible because the XML format which contains the business process description is enhanced by information that covers graphical aspects like symbols, the position of symbols, lines etc..

In \cite{KurzExchangeStandard2016} the practical exchange of BPMN models between different tools is considered in more detail and in \cite{BPMN-Model-Interchange} a demo is shown. In this demo, several interchange scenarios are investigated.

This demo also shows that the interchange of models between various modelling tools does not work without difficulties (therefore this BPMN model interchange working group exists). The interchange of models consisting of several pools with message exchange is not investigated in these demos \cite{BPMN-Model-Interchange} but for distributed business processes this aspect is essential.

If graphical modelling languages are applied the graphical representation of the symbols and their positions on a drawing represents a process model. The BPMN standard includes a language which allows to describe the graphical representation of models.

This means with some restrictions BPMN models can be handled by different modelling tools.

\subsection{Is there a storage language for which formal execution semantics are defined?}

The authors could not find a storage format that can be interpreted by various workflow engines except for BPMN.
In \cite{KossakIGKNZKFS14} the problems with the XML notation of BPMN and the related semantics is discussed in detail.

For the \acrshort{BPMN} storage format a semantic specification in natural language exists. This semantic specification contains “inconsistencies between descriptions of the same element, while at the same time, the semantics of certain elements remains ambiguous” \cite{KossakIGKNZKFS14}.  Nevertheless, subsets of the \acrshort{BPMN} language can be directly interpreted by workflow engines. A list of \acrshort{BPMN} based workflow engines can be found in Wikipedia \cite{WikipediaBPMNengines}. These subsets mainly focus on the control flow elements like action, xor etc..
In \cite{KossakIGKNZKFS14}, for a great part of \acrshort{BPMN} elements a formal semantic has been created. But in this work, pools and swim lanes which are part of \acrshort{BPMN} are not considered. In the \acrshort{BPMN} standard their usage is not specified (see page 306 in \cite{BPMN-OMG-2012}). Especially for business processes in networked organisations, modelling elements that can be used to express a network of organisations are needed. This means that there are problems in modelling processes for networked organisations.\\
The BPMN language suffers from a lack of formalisation. This implies that the \acrshort{BPMN} semantics is not precisely enough that it can be directly executed by a computer.

\subsection{Does a workflow engine exist which is able to execute a process logic available in a specific storage format that has been produced by different methods and tools?}

Currently, a storage format that is used by several modelling methods and related tools could not be found. This also applies to a workflow engine which understands process models created either by BPMN tools or EPK-based modelling tools.

Since the BPMN storage format is standardised, different workflow engines can understand BPMN-based process models created by different modelling tools. 
If models adhere to this standard, in principle, different workflow engines should understand \acrshort{BPMN} models. However, because of some imprecise definitions in the BPMN standard, some problems may arise (see \cite{BPMN-Model-Interchange}) in the practical work.

\section{Structure of the feasibility study}
\label{section:structure-feasability}

\begin{figure}[htb]
    \includegraphics[width=0.485\textwidth]{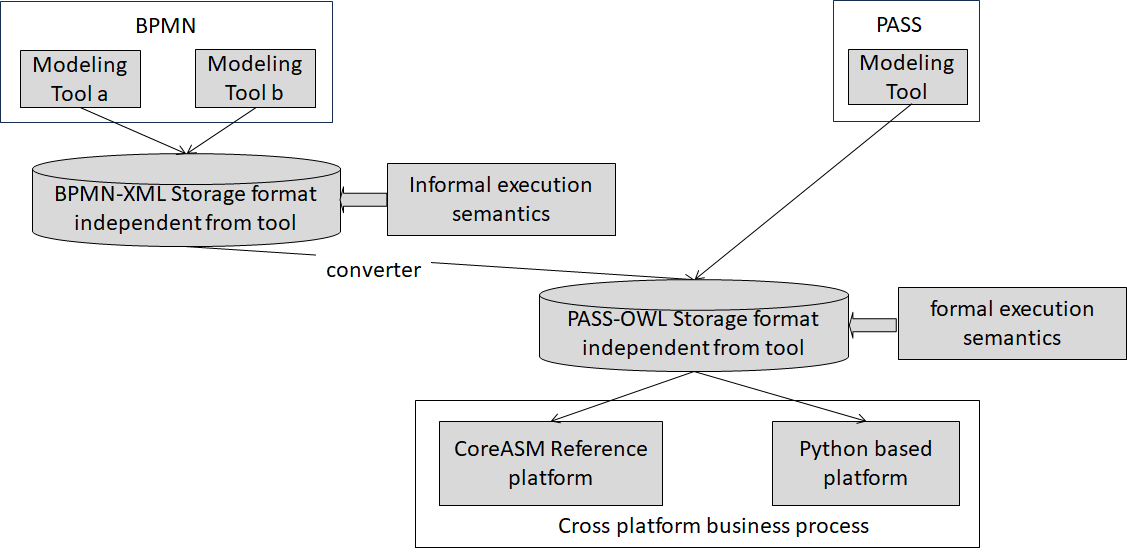}
    \centering
    \caption{Aspects considered in feasibility study}
    \label{fig:Aspects-Feasability-Study}
\end{figure}

 \autoref{fig:businessProcessDigitalisation} shows the aspects which have to be considered if several different modelling tools and execution platforms are used for the digitisation of a business process. In this feasibility study, the combined use of PASS- and BPMN-based modelling tools and execution platforms is considered as a subset of those aspects. \autoref{fig:Aspects-Feasability-Study} depicts the aspects considered in this feasibility study.

The subject-oriented specification language PASS \cite{fleischmannPrimerSubjectOrientedBusiness2012} has been chosen due to its OWL storage format for which a formal execution semantic has been defined \cite{fleischmannSubjectOrientedBusinessProcess2012a}. For the definition of the execution semantics, an \acrfull{ASM} \cite{fleischmannSubjectOrientedBusinessProcess2012a} has been created. Because the PASS language consists only of 5 basic symbols the execution semantics definition does not need to be extensive (about 3 pages).
For \acrshort{ASM} exists an interpreter called coreASM \cite{coreASM} which can execute the OWL storage format of PASS-based models \cite{Logic-Computation}. The disadvantage of PASS is that it is not commonly used. 

BPMN is widely used in the industry since it is an OMG standard and there are many business people trained in that modelling method. The complete BPMN language consists of around 160 different symbols, but in practice around 15 are used.

In \cite{article:S-BPM-vs-BPMN-empirical-evaluation} PASS and BPMN have been compared with each other and it shows up that at least for beginners PASS could be more easily learned than BPMN. For BPMN there exist many modelling tools and execution platforms but these execution platforms understand only the BPMN-XML storage format. Due to the missing formal semantics, different workflow engine may interpret the same process model differently.

The feasibility study described in this paper aims to combine PASS aspects with BPMN advantages. BPMN language features are mapped to PASS language features. Which results in a formal semantic for the corresponding BPMN language features. This means that the formal definitions and simplicity of PASS is combined with the widely known and used modelling language BPMN.
\section{Outline of PASS}
\label{section:PASS}

The \acrfull{PASS} follows the paradigm of subject-orientation when describing models of processes or rather process systems. In such models the term "Subject" refers to active entities. They execute operations on objects and exchange data via messages to synchronise their operations. 
In the following, an informal overview is presented . A precise and formal definition of PASS can be found in \cite{fleischmannSubjectOrientedBusinessProcess2012a}.

\subsection{General concept of subject-orientation}

\begin{figure}[htb]
    \includegraphics[width=0.485\textwidth] 
    {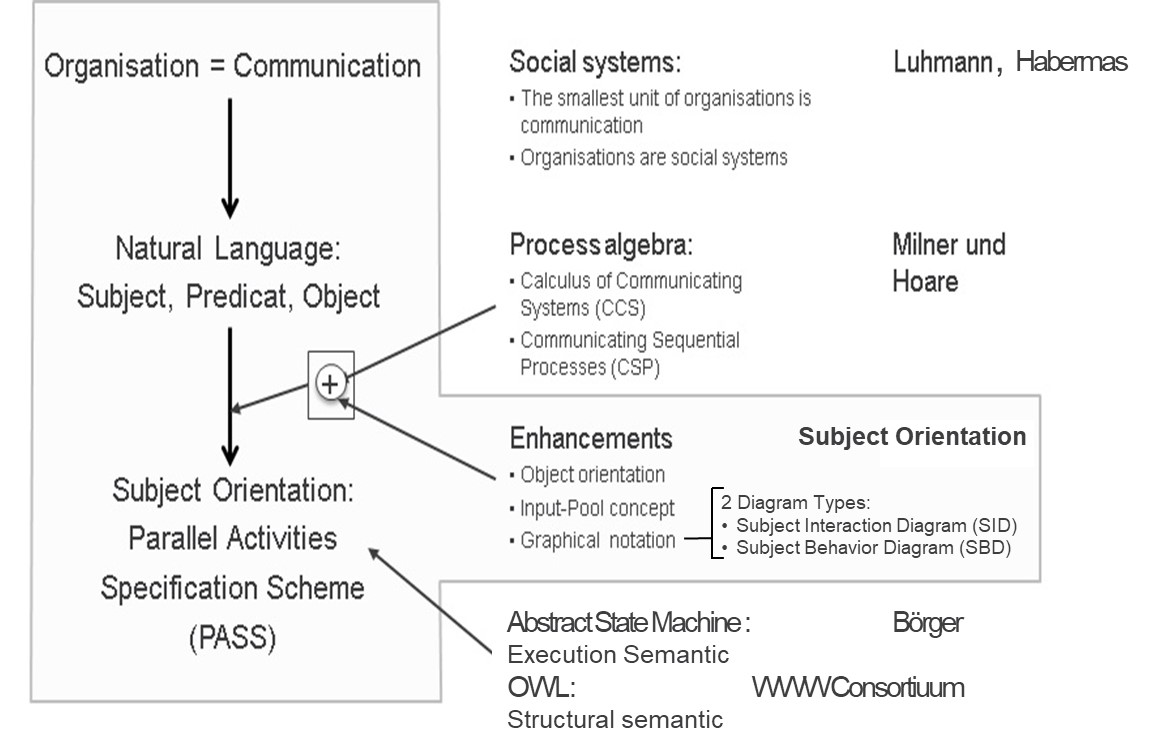}
    \centering
    \caption{Fundamentals of subject-orientation}
    \label{fig:Fundamentals of Subject-Orientation}
\end{figure}

Subject-oriented system development has been inspired by various process algebras (see e.g. \cite{fleischmannSubjectOrientedBusinessProcess2012a}, \cite{book:Luhmann}, \cite{book:CCS}. \cite{book:CSP}), by the basic structure of nearly all natural languages (subject, predicate, object) and the
systemic sociology developed by Niklas Luhmann \cite{book:Luhmann} and Jürgen Habermas \cite{book:Habermas}. In the active voice of many natural languages, a complete sentence
consists of the basic components subject, predicate and object. The subject represents the active element, the predicate (or verb) the action, and the object is the entity on which the action is executed. According to the organisational theory developed by Luhmann and Habermas, the smallest organisation consists of communication executed between at least two information processing entities (Note, this is a definition by a sociologist, not by a computer scientist). \autoref{fig:Fundamentals of Subject-Orientation} summarises the different inspirations of subject-orientation.

Particularly, it describes how these various ingredients are combined in an orthogonal way to a modelling language for scenarios in which active entities play a prominent role like in Industry 4.0.
\subsection{Structure of models described in PASS}
\begin{figure}[htb]
    \includegraphics[angle=90, width=0.485\textwidth]{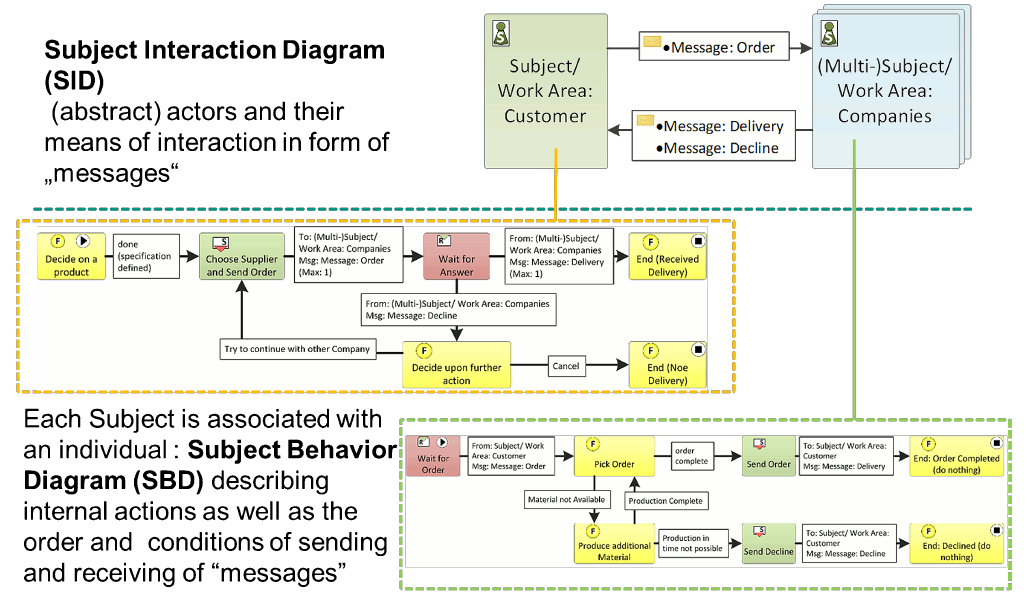}
    \centering
    \caption{Main components of PASS}
    \label{fig:Main Components of PASS}
\end{figure}

A \acrshort{PASS} model (system) consists of two separate but interconnected graph descriptions. First, the \acrfull{SID} that denotes the existence of active entities (the subjects) and their communication relationships, that they can use to exchange data in a process context. Furthermore, for each subject there can be an individual \acrfull{SBD} that denotes its specific activities in a process. A subject acts upon (data) objects that are owned by the subject and can not be seen by other subjects. It is important to emphasise, that this specification is totally independent from the implementation of subjects, objects, and the communication between subjects. This means subjects are abstract entities which communicate with
each other and use their objects independently from possible implementations. The mapping to an actual implementation entity or technology is done in a succeeding step. When an implementation technology is assigned (mapped) to a subject to execute its behaviour (\acrshort{SBD}) it becomes an actor/agent, e.g. a software agent. In \autoref{fig:Main Components of PASS} a \acrshort{PASS} model is shown. The upper part of that figure shows the graphical representation of a \acrshort{PASS} model with \acrshort{SID} (upper diagram) and SBDs (lower diagrams). In the example subject 'Customer' sends a message 'Order' to the subject 'Companies' and receives the messages 'Delivery' or 'Decline'.

In principle, an \acrshort{SBD} defines the behaviour of a subject as a kind of state machine, interlinking three types of states (see the lower part of \autoref{fig:Main Components of PASS}). Send-states (green nodes) represent the dispatch of messages to other subjects, receive-states (red nodes) represent the reception of messages from other subjects, and function/do-states (yellow nodes) represent tasks that do not involve interaction
with other subjects. States are connected using transitions representing their sequencing. The behaviour of a subject may include multiple alternative paths. Branching in PASS is represented using multiple outgoing transitions of a state,
each of which is labelled with a separate condition. Merging of alternative paths is represented using multiple incoming transitions of a state. Within an \acrshort{SBD}, all splits and merges in the process are always explicitly of the XOR type.
There are no AND or OR splits! Subjects are executed concurrently. Triggering and synchronising concurrent
behaviours is handled by the exchange of messages between the respective subjects. For a subject-oriented process model to be complete and syntactically correct, all messages specified in the SID (and only those) must be handled in the
SBDs of the two subjects involved. The \acrshort{SBD} of the sending subject needs to include a send-state specifying the message and recipient name (see \autoref{fig:Main Components of PASS}). Correspondingly, the \acrshort{SBD} of the receiving subject needs to include a receive-state specifying the message and sender name. There is no explicit diagrammatic association of the messages shown in the SID with the corresponding send and receive-states in the SBDs. At runtime, any incoming message is placed in the so-called input-pool of the
receiving subject, which can be thought of as a mailbox. When the execution of the subject has reached a receive-state that matches the name and sender of a message in the input-pool, that message can be taken out of the input-pool and behaviour execution can proceed as defined in the \acrshort{SBD}. The default communication mode is asynchronous. Synchronous communication can be established by restricting the maximum number of messages that can be stored in the input-pool. The input-pool is not visualised in a diagram but is an important concept
in order to understand how messages and behaviours are loosely coupled with subjects in PASS.
\section{Outline of BPMN}
\label{section:BPMN}

BPMN – the modelling language referred to as Business Process Modelling (and) Notation – was developed by IBM in 2002 and subsequently published by the BPMI (Business Process Management Initiative). The aim was to create a universally applicable standard to counter the multitude of process modelling languages used in academia and industry. This language should adopt the essential characteristics of the most common languages and make it possible, in addition to the documentation of business processes, to create models that allow for immediate IT-supported execution. BPMI in turn was merged with the OMG (Object Management Group) in 2005. Thus, BPMN became an OMG standard. 
The BPMN 2.0 standard was published in 2010. This standard incorporates several diagram types: the choreography diagram, the conversation diagram, and the collaboration diagram. In the following, the basic elements of BPMN 2.0 that enable business processes to be represented at the business level are described.
BPMN focuses on business processes, which it presents as a temporally logical sequence of activities (tasks) that are structured in accordance with organisational responsibilities. The representation of data is not as comprehensive as in other modelling languages and only regarded in the context of process flows.
\\
The BPMN language consists of more than 160 different symbols. Many of these symbols focus on the description of control flows (similar to flowcharts, EPCs, or activity diagrams), but in this paper, mainly the processes running across different organisations are considered. This means communication aspects are more relevant. Therefore there is a focus mainly on these modelling elements.

\subsection{Notation elements for modelling process flows}
\begin{figure}[htb]
    \includegraphics[width=0.485\textwidth]{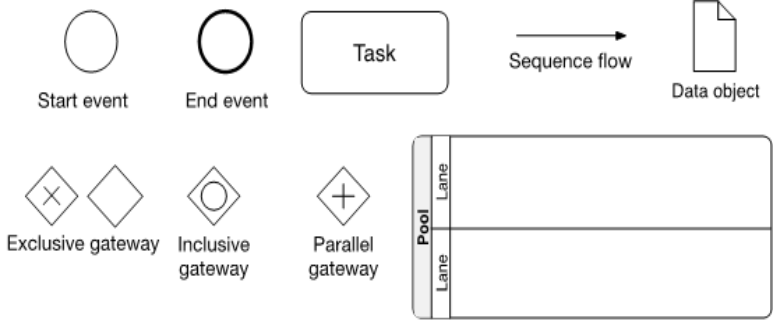}
    \centering
    \caption{Core notation elements of BPMN}
    \label{fig:Core notation elements of BPMN}
\end{figure}

Process diagrams created with BPMN are called Business Process Diagrams (BPD). At its core a BPD follows the principles of activity diagrams, which are subsequently supplemented by elements that allow the representation of the potentially more complex control flow in business processes (see \autoref{fig:Core notation elements of BPMN}).

As a basic principle of representation, certain things have to be done in a process (tasks), but possibly only under certain conditions (gateways), and things can happen (events). These three objects are connected to each other via sequence flows, however only within the confinements of pools or lanes. Pools and lanes are constructs used to represent responsibilities in distributed business processes. They are discussed in more detail below. If a connection is made across pool boundaries, it is modelled using message flows.
\\
A process consists of tasks. After starting a process (by means of an event), one task follows the other until the process ends (with an event). Tasks can be atomic (i.e. not refined further) or can contain sub-processes. In such cases, tasks are refined by an additional embedded BPD, which represents its detailed sequence of sub-tasks. This detailed sequence can be "hidden" and is represented by a "+" symbol at the bottom of the task.
\\
A process begins with a start event and ends with an end event. BPMN offers a multitude of possibilities to define events that can trigger, complete, or influence the course of a process. These will be discussed later.
\\
At this point, it is important to emphasise that a process can start with one or more start events and can end on any path through the process (see sequence flow and gateways below) with one or more end events. There must be a continuous sequence flow from each start event to at least one end event. Tasks, gateways, or intermediate events must not be endpoints in the process and therefore always require at least one outgoing sequence flow.
A gateway represents a branch in the control flow. The exclusive (XOR) gateway requires a condition for each outgoing control flow, which according to the standard must always refer to the result of an immediately preceding task. 
\\
The parallel (AND) gateway tracks all outgoing control flows independently and in parallel. The branched control flows can be terminated separately with end events or explicitly merged again with another parallel gateway. After this merger, the control flow only continues once all incoming control flows have been completed (as with the split/join concept for activity diagrams).
The inclusive (OR) gateway can follow one or more paths, whereby a condition must be specified for path selection (as with the exclusive gateway). This condition must already be testable at the time of the decision, so the necessary data must have been generated in one of the previous tasks.
Decisions, which cannot be made on the basis of previously existing data, can be represented using the event based gateway. This requires an event in each outgoing branch immediately after the gateway (e.g., an incoming message event or a timer event). When one of these events occur, its respective branch (and only this branch) is activated.

\subsection{Notation elements for modelling communication}

\begin{figure}[htb]
    \includegraphics[width=0.485\textwidth]{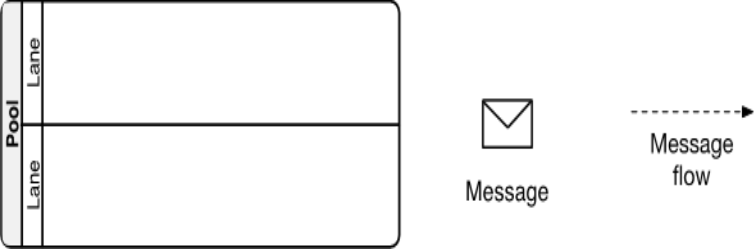}
    \centering
    \caption{BPMN notation elements for modelling communication}
    \label{fig:BPMN notation elements for modeling communication}
\end{figure}

BPMN also enables the modelling of distributed business processes. Although BPMN clearly focuses on the process flow during modelling (similar to flowcharts, EPCs, or activity diagrams), it also enables the structuring of the process in accordance with the participants involved and their associated responsibilities. The modelling elements available for this purpose are described in this section (see \autoref{fig:BPMN notation elements for modeling communication}).

A pool represents a company or an organisational unit in a company, such as a department. Each (swim) lane in a pool represents a person or role involved in the process that is assigned to this pool.
\\
BPMN allows the representation of the interaction of two or more processes. The aforementioned pools and lanes are necessary for the representation of collaborations. Separate lanes are required for all persons or groups involved in a process, and separate pools are necessary for each process or organisational unit that is responsible for this process. Each pool thus contains its distinct processes with separate start and end events. Nevertheless, these individual processes can strongly influence each other, in which case they are coupled via message flows.
\\
Message flows indicate that data is exchanged between different processes. Therefore, no message flow can take place within a process (pool). Consequently, there are no message flows within a lane or between different lanes of a single pool. Sequence flows show which activities are executed in which order and do not explicitly constitute an exchange of data. In contrast to message flows, they may only be used within a pool and not between different processes (pools).
\\
Message flows can be augmented with message elements, which are used to explicitly represent the exchanged data and contain a more precise specification of the transmitted information.
Message flows can be used in different ways: they may either originate from pools and activities and also end there, or they can be explicitly sent by send message events and received by receive message events. The first case is useful for the descriptive modelling of business processes in which a communication process is to be represented that does not necessarily have to be described exactly. A message originating from an activity or pool is sent at some point during task or process execution - the exact time remains unclear. A message ending at a pool only states that the represented organisation receives this message, but not which activity it triggers or how it is handled within a process. This can be useful when modelling external organisational units, whose detailed behaviour is unknown. An exact specification of communication processes, however, is only possible by using explicit send and receive events.

\begin{figure}[htb]
    \includegraphics[width=0.485\textwidth]{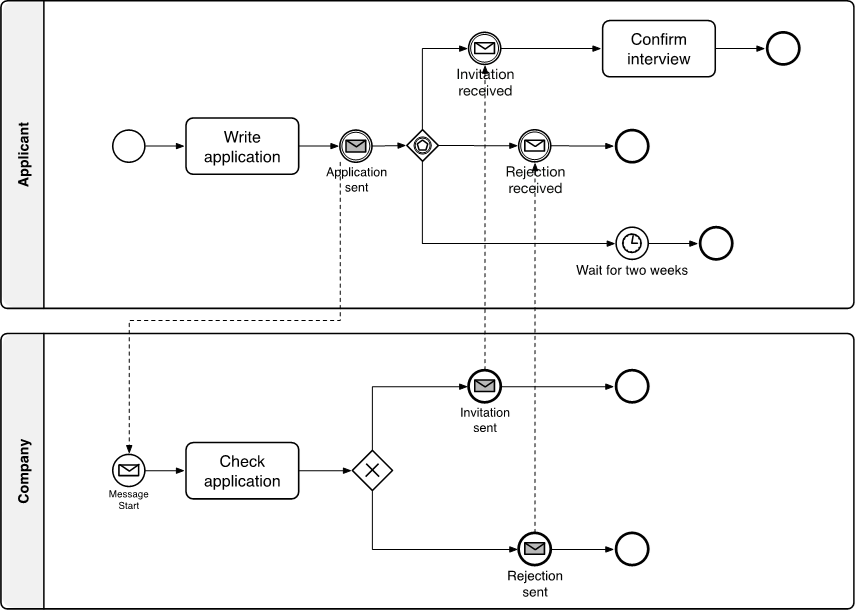}
    \centering
    \caption{Example BPMN diagram showing communication-oriented processes}
    \label{fig:Example BPMN diagram showing communication-oriented processes}
\end{figure}

The example in \autoref{fig:Example BPMN diagram showing communication-oriented processes} shows two processes (one per pool) that are linked by message flows. The company's process is triggered by an incoming application, which is represented here in the first message flow. After checking the application, the decision can be made whether to send an invitation or reject the application. In the upper pool, the applicant waits for an answer for a maximum of two weeks (as represented by the intermediate timer event). The event based gateway activates the process branch whose event occurs first. The related send and receive events are linked via message flows. It is important to note here that message flows always represent 1:1 relationships – that is, a sent message can be received exactly once and a receive event can react to exactly one message. This is a significant difference between PASS and BPMN. In PASS messages are exchanged via a so-called input-pool in which messages sent to the related subject are deposited. This means that sending states in the sending process does not need 
a one-to-one relationship to the receiving state in the receiving subject (see \autoref{section:PASS}). In order to emulate that PASS feature in BPMN signal events are required which are not covered in the BPMN conversion (see \autoref{tab:mapping}).

\section{Transformation of BPMN models}
\label{section:Transformation}

This section describes how \acrshort{BPMN} processes are converted into \acrshort{PASS} processes.
This is done in two steps whereas the XML-format which is used for storing BPMN models is transformed into the OWL-format used for storing PASS models.

First, the \acrshort{BPMN} model is converted into a SimpleBPMN model.
Unnecessary information and elements are filtered out.
In addition, some information is added to the elements in order to simplify the subsequent conversion into a \acrshort{PASS} model.
The SimpleBPMN elements are then converted into \acrshort{PASS} elements and exported in OWL-format.

The process as a whole is not translated in the respective steps, but each element is translated separately.
At the end, the individual elements are reassembled into a fully translated process.
This approach breaks down the complex problem into several individual parts that can be solved individually.

\autoref{tab:mapping} shows an overview of which \acrshort{BPMN} elements are translated into which \acrshort{PASS} elements.

\begin{table}[htb]
	\caption{Mapping of BPMN to PASS elements}
	\label{tab:mapping}
	\centering
	\begin{tabular}{|p{0.15\textwidth}|p{0.29\textwidth}|}
		\hline
        \textbf{BPMN element}                 & \textbf{PASS element} \\
        \hline
        \multirow{1}{*}{Participant}            & Subject \\
        \hline
        \multirow{3}{*}{Message flow}            & Message specification \\
                                                & Message exchange \\
                                                & Message exchange list \\
        \hline
        \multirow{1}{*}{Process}                & Subject behaviour diagram \\
        \hline
        \multirow{2}{*}{Start event}            & Do-state (Initial state of behaviour) \\
                                                & Action \\
        \hline
        \multirow{4}{*}{Message start event}    & Receive-state (Initial state of behaviour) \\
                                                & Action \\
                                                & Receive transition \\
                                                & Receive-transition-condition \\
        \hline
        \multirow{2}{*}{End event}              & Do-state (End-state) \\
                                                & Action \\
        \hline
        \multirow{2}{*}{Task}                   & Do-state \\
                                                & Action \\
        \hline
        \multirow{2}{*}{Exclusive gateway}      & Do-state \\
                                                & Action \\
        \hline
                                                & Send state \\
        Intermediate throw                      & Action \\
        \hspace*{0.15cm} message event          & Send-transition \\
                                                & Send-transition-condition \\
        \hline
        \multirow{2}{*}{Event based gateway}    & Receive-state \\
                                                & Action \\
        \hline
                                                & Receive-state \\
        Intermediate catch                      & Action \\
        \hspace*{0.15cm} message event          & Receive-transition \\
                                                & Receive-transition-condition \\
        \hline
                                                & Do-state \\
        Intermediate catch                      & Action \\
        \hspace*{0.35cm} time event             & Day-time-timer-transition \\
                                                & Day-time-timer-transition-condition \\
        \hline
        Sequence flow                           & Do-transition \\
        \hline        
	\end{tabular}
\end{table}

\subsection{Translation of the individual elements}

This section describes how \acrshort{XML} \acrshort{BPMN} elements are converted into \acrshort{OWL} \acrshort{PASS} elements.
Each element in a \acrshort{BPMN} process has a unique ID.
This is used to describe the connections between the individual elements.
During translation, the \acrshort{BPMN} IDs become the \texttt{ComponentId} attributes that each \acrshort{PASS} element needs.
This allows the elements to be reassembled into a complete \acrshort{PASS} model after translation.
Another advantage of this approach, where the \acrshort{BPMN} IDs are used for the \acrshort{PASS} \texttt{ComponentId} attributes, is that when translating back, which must be done to validate the results, the \acrshort{BPMN} IDs can be restored.

The names of the \acrshort{BPMN} elements are used to generate the \texttt{ComponentLabel} attributes of the \acrshort{PASS} elements.

Each state in a \acrshort{PASS} model has an associated action element.
The state and all associated transactions are stored in this element.
When a state is generated during compilation, the associated action element is also generated.
Also, all transactions are added to the correct action element when created.

Each transition in a \acrshort{PASS} model serves as a connection between two states and therefore possesses a \texttt{sourceState} element and a \texttt{targetState} element.

In the following subsections, each element shown in the table is described briefly.
It also briefly shows what had to be taken into account when implementing the translation software.

\subsubsection{Participants}

One of the most important elements of a \acrshort{PASS} process are subjects.
These are generated from the \texttt{Participant} elements from \acrshort{BPMN}.

Each \texttt{participant} is converted to a \texttt{subject}.

\subsubsection{Message flow}

Besides the subjects, the message flow is one of the main components of a \acrshort{PASS} process.
This is generated from the \texttt{MessageFlow} elements.

Each \texttt{Message\linebreak[0]Flow} is converted into a \texttt{Message\linebreak[0]Speci\linebreak[0]fication} and a \texttt{MessageExchange}.
These are then stored in a \texttt{Message\linebreak[0]Exchange\linebreak[0]List}.

In the \texttt{Message\allowbreak Specification} the content of the message is stored.
In the \texttt{MessageExchange} the corresponding \texttt{sender} and \texttt{receiver} \texttt{subject} are stored.
All this information is read from the \texttt{MessageFlow} element.

\subsubsection{Process}

Each subject in a \acrshort{PASS} process has a \acrshort{SBD} that is stored as \texttt{SubjectBehavior}.
These elements are generated from the \texttt{Process} elements.

Each \texttt{Process} is converted to a \texttt{SubjectBehavior}.

\subsubsection{Start event and message start event}

In a \acrshort{PASS} process there is always a state which is an initial state of behaviour.
This is created from the \texttt{Start\allowbreak Event} or the \texttt{Message\linebreak[0]Start\linebreak[0]Event} element.

A \texttt{Start\linebreak[0]Event} is converted the same as a \texttt{Task} element.
A \texttt{Message\linebreak[0]Start\linebreak[0]Event} is converted the same as a \texttt{Intermediate\linebreak[0]Catch\linebreak[0]MessageEvent} element.
In both cases the \texttt{Initial\linebreak[0]State\linebreak[0]Of\linebreak[0]Behavior} attribute is set for the state.

\subsubsection{End events}

In a \acrshort{SBD} of a \acrshort{PASS} process there are always one or more states which are also end states.
These are created from the \texttt{EndEvent} elements.

An \texttt{EndEvent} is converted the same as a \texttt{Task} element.
After that, the \texttt{EndState} attribute is set at the state.

\subsubsection{Task and exclusive gateway}

In a \acrshort{SBD} of a \acrshort{PASS} process there are usually one or more do-states.
These elements are created from the \texttt{Task} and \texttt{ExclusiveGateway} elements.

\subsubsection{Intermediate throw message event}

To send messages to other subjects, there are one or more send-states and send-transitions in a \acrshort{PASS} process.
These elements are created from the \texttt{Intermediate\allowbreak Throw\allowbreak Message\allowbreak Event} elements.

Each \texttt{Inter\linebreak[0]mediate\linebreak[0]Throw\linebreak[0]Message\linebreak[0]Event} is converted to a \texttt{Send\linebreak[0]State}, a \texttt{Send\linebreak[0]Transition} and a \texttt{Send\linebreak[0]Transition\linebreak[0]Condition}.

The \texttt{SendTransitionCondition} contains the corresponding \texttt{MessageExchange} and the \texttt{messageSentTo} attribute.

\subsubsection{Event based gateway}

To wait for messages from other subjects, there are one or more receive-states in a \acrshort{SBD} of a \acrshort{PASS} process.
These elements are created from the \texttt{EventBasedGateway} elements, among others.

\subsubsection{Intermediate catch message event}

To receive messages from other subjects, there are one or more receive-states and receive-transitions in a \acrshort{PASS} process.
These elements are created from the \texttt{Intermediate\linebreak[0]Catch\linebreak[0]Message\linebreak[0]Event} elements.

Each \texttt{Intermediate\linebreak[0]Catch\linebreak[0]Message\linebreak[0]Event} is converted to a \texttt{Receive\linebreak[0]State}, a \texttt{Receive\linebreak[0]Transition} and a \texttt{Receive\linebreak[0]Transition\linebreak[0]Condition}.

If the \texttt{IntermediateCatchMessageEvent} is after a \texttt{EventBasedGateway}, the \texttt{ReceiveTransitions} will be attached directly to the \texttt{ReceiveState} created from it.
Otherwise, a separate \texttt{ReceiveState} must be created.

The \texttt{Receive\linebreak[0]Transition\linebreak[0]Condition} contains the corresponding \texttt{Message\linebreak[0]Exchange} and the \texttt{message\linebreak[0]Sent\linebreak[0]From} attribute.

\subsubsection{Intermediate catch time event}

To measure time and thereby trigger state transitions, there are one or more day-time-timer-transitions in a \acrshort{PASS} process.
These elements are created from the \texttt{Intermediate\linebreak[0]Catch\linebreak[0]Time\linebreak[0]Event} elements.

Each \texttt{Intermediate\linebreak[0]Catch\linebreak[0]Time\linebreak[0]Event} is converted to a \texttt{Day\linebreak[0]Time\linebreak[0]Timer\linebreak[0]Transition} and a \texttt{Day\linebreak[0]Time\linebreak[0]Timer\linebreak[0]Transition\linebreak[0]Condition}.

If the \texttt{IntermediateCatchTimeEvent} is positioned after an \texttt{EventBasedGateway}, the \texttt{Day\linebreak[0]Time\linebreak[0]Timer\linebreak[0]Transition} is attached directly to the \texttt{Receive\linebreak[0]State} created from it.
Otherwise, a separate \texttt{Do\linebreak[0]State} must be created.

The \texttt{DayTimeTimerTransitionCondition} contains the time duration after which the transition is triggered.

\subsubsection{Sequence flow}

Transitions are required between all states in a \acrshort{PASS} process.
So far, all transitions have been created except for the do-transitions.
The missing do-transitions are created from the \texttt{SequenceFlow} elements.

\subsection{Different concepts}

In this work, a major design difference in modelling became apparent.
In the \acrshort{BPMN} standard, modelling is done with tasks.
In the \acrshort{PASS} standard, modelling is done with states.

This makes the translation of \acrshort{BPMN} to \acrshort{PASS} models difficult.
Even though the use of tasks and states works very similarly in the respective standards and it is therefore reasonable to translate tasks into states, this is not entirely correct.

One way around this problem is to model and name the tasks like states and the sequence flows like transitions.
This way the translated \acrshort{PASS} model would be named and modelled correctly.

However, if the model is to be completely correct, it would have to be modelled with intermediate events.

\begin{figure}[htb]
    \includegraphics[width=0.485\textwidth]{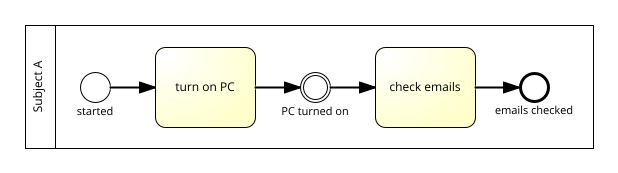}
    \centering
    \caption{Subject-oriented BPMN process with intermediate events}
    \label{fig:ProcZwi}
\end{figure}

This allows the software to convert \acrshort{BPMN} intermediate events into \acrshort{PASS} states and \acrshort{BPMN} tasks into \acrshort{PASS}-transitions.
\autoref{fig:ProcZwi} shows this type of modelling.

Although this would be the correct modelling, for this work it was decided not to support it.
First, this modelling style is not intuitive and also unnecessarily complicated for modellers.
Furthermore, while this is possible with simple processes, as more elements are modelled to it, the translation also becomes more difficult.

In this paper, it was therefore decided that the tasks would be translated into states and the sequence flows into transitions.
In a subsequent project, however, the translation software could be adapted to this type of modelling.
\section{Workflow engine concept}
\label{section: workflow-concept}
In this section, a concept for a code-generating approach for a \acrshort{PASS} workflow engine is described. This workflow engine aims to fulfill the set of requirements for workflow engines discussed in \autoref{requirements}. In \autoref{High_level_view}, the developed concept is introduced. Individual components such as actors and the user interface are discussed in \autoref{internal_actor_concept} and \autoref{user_interface}. The overall goal of this concept is to determine if the approach of generating code from models is feasible in workflow engines while also fulfilling the requirements for such systems. As a result, the workflow engine can serve as a validation tool for the execution logic of the input models as well as the \acrshort{PASS} standard. Especially when compared to an independently developed alternative workflow engine. For the prototype concept and implementation, the underlying conceptual similarities of \acrshort{PASS} to the actor model in computer science are utilised. Furthermore, the implications for a potential multi-enterprise scenario are discussed briefly in \autoref{multienterprise}.

\subsection{Requirements}
\label{requirements}
In \cite{fleischmannPrimerSubjectOrientedBusiness2012} and \cite{singerAgentBasedBusinessProcess2016} certain requirements for software concerning the execution of workflows are defined. Those are the basis on which the concept is developed.

\begin{itemize}
	\item The ability to interpret or execute the process models.
	\item A user interface including a task list, as well as forms with editable and read-only data. Either manually generated or based on the business objects.
	\item Storage for process models that are uploaded via the interface.
	\item Storage for business objects and a wide support of data types.
	\item Storage of information on which operations need to be performed.
	\item A way of integrating other systems (e.g. \acrfull{API} calls).
	\item A method for mapping abstract roles or subjects to organisational structures (e.g. \acrfull{AD} integration).
	\item A permission system based on organisational structures.
	\item Offer users a way to maintain a task list.
	\item The possibility of message exchange.
	\item Monitoring aspects as well as log file generation to allow traceability.
	\item Scalability and ability to handle a large number of process instances.
	\item Persistence in case of restarts.
\end{itemize}

As defined as the goal of this paper, a main requirement is the transformation of models into executable code. Therefore, a code generator component is required to perform this translation.

\subsection{Overview}
\label{High_level_view}
This workflow engine concept aims to be the basis of a workflow engine which fulfils the requirements defined in \autoref{requirements}. Those are requirement for the environment the process lives in and not for the process itself. Therefore, the main component are actor systems containing the actors. Those systems are accompanied by external components for enhanced functionality. Within this concept, the actor system can be seen as a separate program which is the underlying platform for the actors to run in. It ensures that actors can communicate with each other and also manages them (e.g. starting, stopping). Furthermore, it is able to accept new actors at runtime, so that it does not need to be restarted in case an actor changes.

As previously mentioned, the actor system is the basis for actors. In the overall concept, actors are generated on the basis of subjects within a \acrshort{PASS} model. The main distinction is that an actor is a program which is somehow linked to an user or group, while the subject is an abstract role in the model. It is up to the execution environment to perform this translation. It is important that the behaviour of the actor code matches the subject behaviour in the model.

\begin{figure}[htb]
	\centering
	\includegraphics[width=0.485\textwidth]{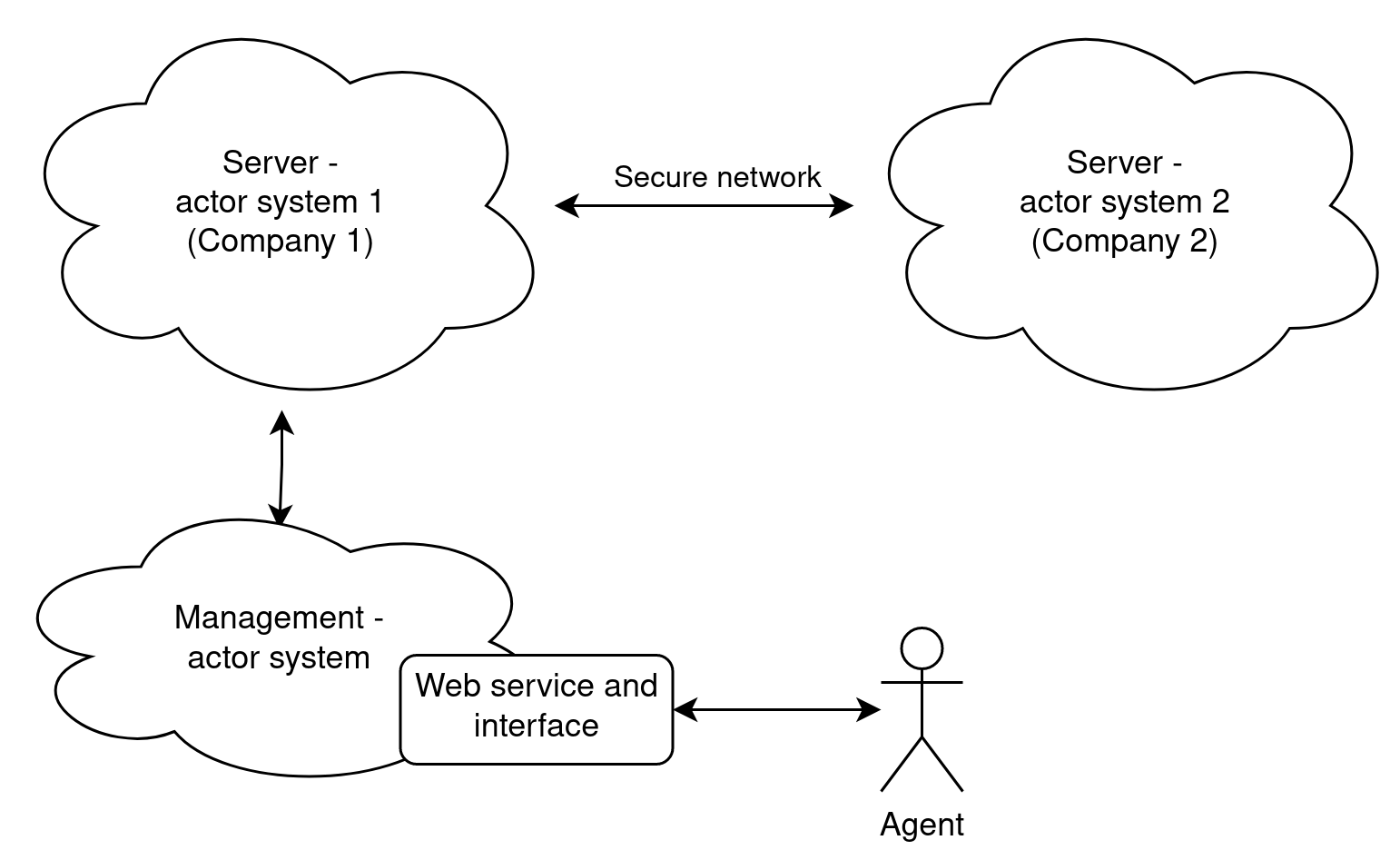}
	\caption{High level view}
	\label{fig:high_level_concept}
\end{figure}

Furthermore, the concept consists of multiple different and potentially distributed interconnected actor systems. This allows actors to communicate with each other even if not in the same actor system (i.e. allowing message exchange). Since those actor systems can be distributed over multiple servers, redundancy, as well as performance, can be increased and scaled. This structure aims to fulfill the requirement of scalability and the ability to deal with a large number of process instances. In addition, this approach potentially allows inter-company execution of processes where multiple actor systems running in different companies are connected. In such a scenario, actors from one company could directly interact with actors from other companies while the actor systems act as a common basis for communication. A high-level view including two companies is depicted in \autoref{fig:high_level_concept}.

On a high level, the basic functionalities are represented using two different kinds of actor systems. While technically both perform the same functionality, they serve a different purpose in the concept. This includes the server-side actor system where the actual business logic (i.e. the actor generated from a subject) is executed as well as the actor system for management. The second is responsible for providing an interface for external tools such as user interfaces. This is done by separate (fixed and not generated) actors providing an entry point for external communication. As shown in \autoref{fig:high_level_concept}, those actor systems could potentially exist in different companies. Given that a network connection exists, they are able to provide a way of communication between the companies.

As previously defined it is a requirement that the users are able to interact with the system (see \autoref{requirements}). For this purpose, a web interface could be used which communicates through the management actor system (see \autoref{fig:high_level_concept}) with the server actor system. As a result, the user interface is separated and can exist on a different physical or virtual computer. However, in the depicted concept, the user interface is tightly coupled to an actor system which allows for utilisation of the internal message exchange capabilities. This provides the benefit that the asynchronous nature of actor systems can also be utilised in the interface and no separate means of communication have to be implemented.

\subsection{Internal actors}
\label{internal_actor_concept}
As previously mentioned, the concept consists of at least two different interconnected actor systems. This interconnection allows communication between actors even if running on other actor systems. Furthermore, the actor system itself is able to communicate with actors and can therefore be used for external communication. In addition, it can also be integrated into other components. This is used for the management actor system included in the web interface where communication occurs between the management actor system and actors running on the server-side actor system.

\begin{figure}[htb]
	\centering
	\includegraphics[width=0.485\textwidth]{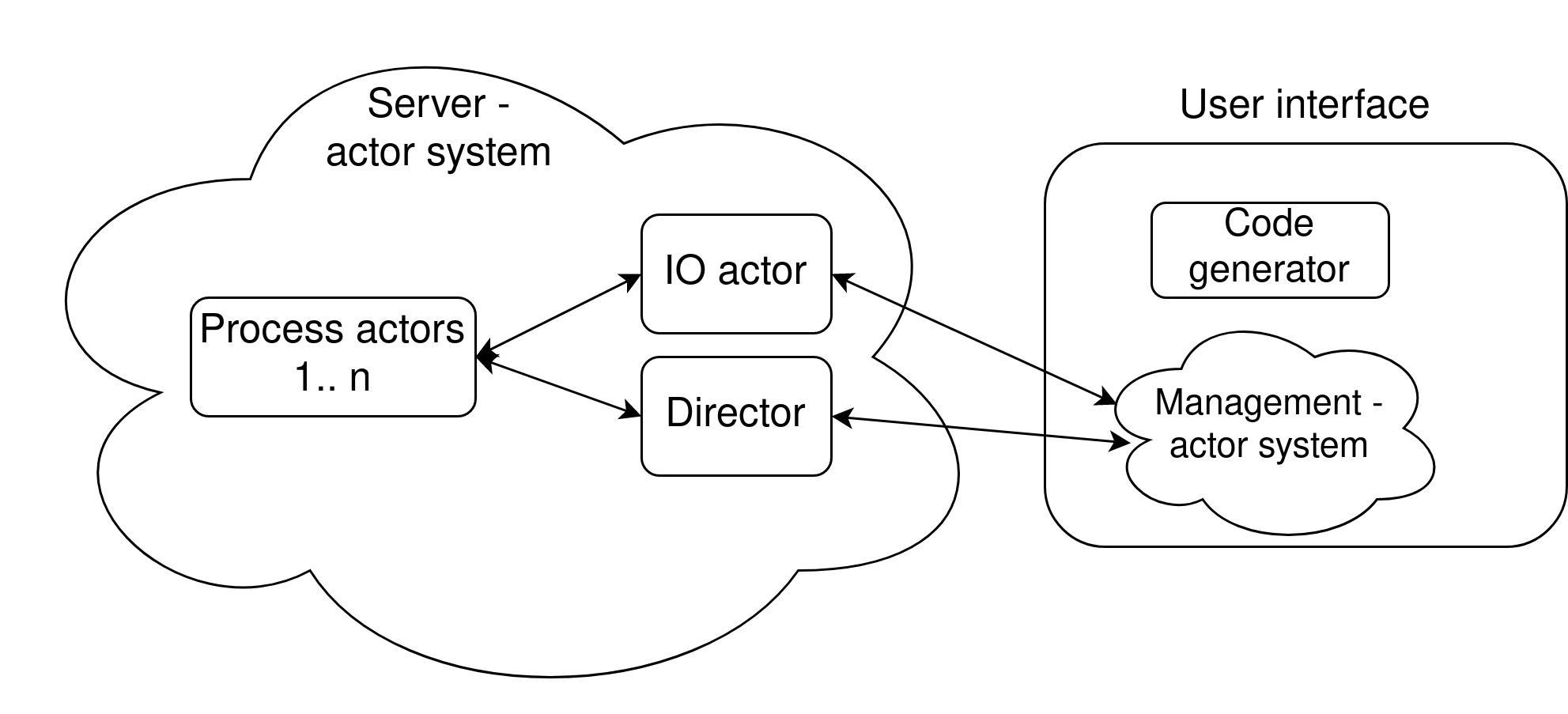}
	\caption{Detailed view on communication paths}
	\label{fig:detail_concept}
\end{figure}

However, since the actor system is volatile (actors can be started or stopped at any time) when it comes to the actor addresses as well as the number of actors running, two fixed actors, namely director and IO actor, are introduced. Those fixed points provide a stable reference for external tools to interact with the system. A detailed view of the structure concerning the communication of those actors is depicted in \autoref{fig:detail_concept} and is discussed in the following section.

\subsubsection{Director}
\label{Director_concept}
Within a process model, the starting subjects (and therefore actors) are defined and started by the system on process instantiation. All subsequent ones are started on demand when a message is sent to them by another actor. As part of the starting process, the address of the newly created actor is known to the sender. Since the sender is also included in the message, the recipient (newly started actor) also knows the address of the sender. Therefore, in the simplest case, no actual actor discovery is needed. 

However, this is only the case as long as only one start actor exists. In addition, the message flow can only occur in the order of actor creation which does not fit real-world applications e.g. processes with multiple start actors or actors communicating with actors not started by them. To prevent falsely starting actors, actor discovery is needed and implemented in the director actor. 

The director has to store a variety of information about the currently running actors. This includes the association of actors to a specific instance to ensure that no messages are sent between instances (resulting in data leakage and potential confusion). To utilise the data of the director, each actor has to perform a lookup of some sort before sending a message, especially creating a new actor. It is then the responsibility of the director to hand out the addresses of the already existing actors. This is also depicted in the detailed communication paths shown in \autoref{fig:detail_concept}.

The main point of the director actor approach is to provide a central point for all actors to resolve the addresses of message targets. This raises the question how the director gets this information. This can be done by various approaches, where the following was chosen for the prototypical implementation.

\begin{figure}[htb]
	\centering
	\includegraphics[width=0.258\textwidth]{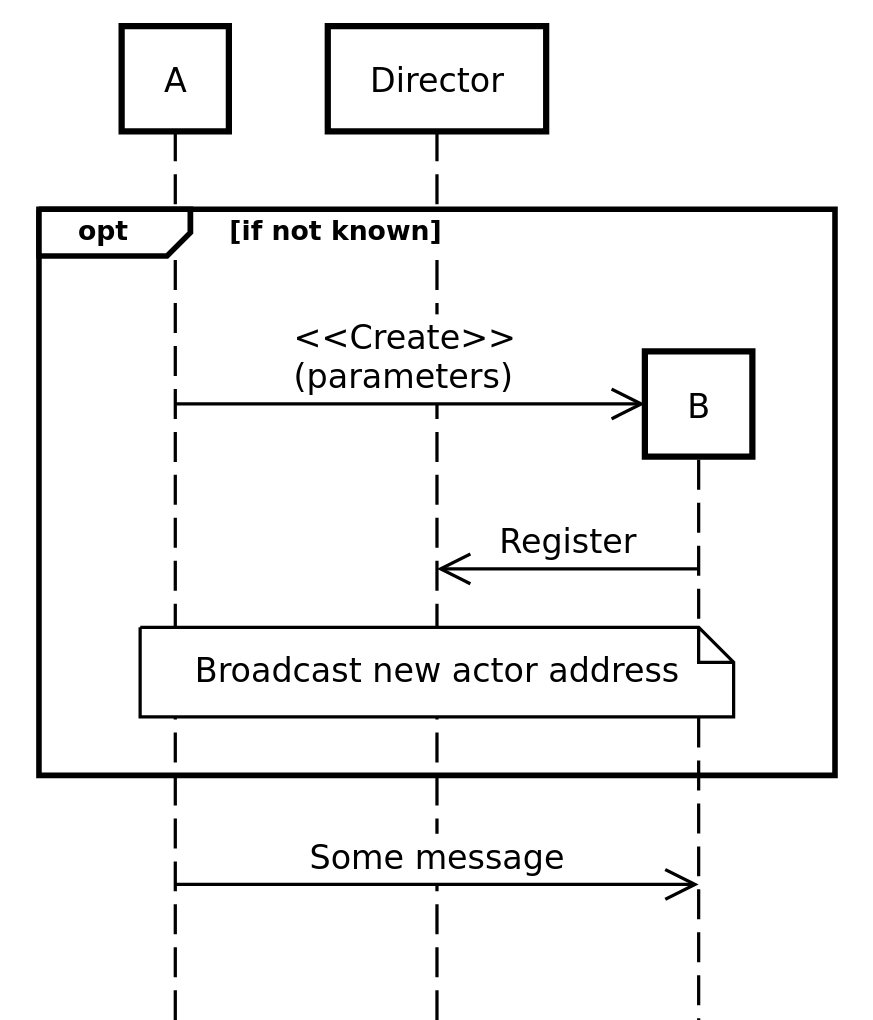}
	\caption{Actor discovery}
	\label{fig:registerapproach2}
\end{figure}

This approach (\autoref{fig:registerapproach2}) places the create action to the actor A if the address of B is not known. If B is created by A, it registers itself with the director actor. The director broadcasts this information to all actors within this instance (including B). Either due to the direct creation process or by the broadcast, A is now aware of B and vice versa. This would also be the case for a fictional actor C which receives the broadcast and stores the addresses locally. Therefore, for a fictional C, the locally stored address would be reused, and no actor creation would be performed. Therefore, it is crucial that newly created actors register themselves with the director immediately after instantiation.

Another task of the director is keeping process instances separated by grouping actors together. For this reason, the director needs some form of ID to identify correlated actors. This ID has to be used by the actors to identify themselves while communicating with the director. Therefore, a parametrisation is necessary at translation and instantiation time. The first includes information about the model, such as the name and the actor system the actor should run at (e.g. the company in multi-company scenarios). The second contains information about the instance, including the instance ID.

As those steps are done in separate components in the concept, the information is also stored in different places. While information at translation time is stored in the user interface component (i.e. database, and also hard-coded in the generated program code), the instance information is only stored in the director actor hence in memory. The latter is also applicable to business objects. As a result, the persistence of the actor system is a factor to be considered. 

The functionality of the director can further include instance management such as starting and stopping actors. Since the director has a known address, it is possible for the interface to communicate with the director through the management actor system. As a result, the functionality of the director can be utilised in the web service as well (e.g. status page, starting or stopping actors).

\subsubsection{IO actor}

As defined in the requirements, a workflow engine has to support user interactions. For this concept, this is defined as an input or output action to the user (i.e. reading or manipulating business objects). In the concept depicted in \autoref{fig:detail_concept} this functionality is designed in two steps. First a separate actor, the IO actor, is used, to utilise the actor system's internal and external communication mechanisms. Second is the user interface which enables the user to communicate with the IO actor. The user interface is discussed in detail in \autoref{user_interface}.

The idea behind the IO actor is similar to the director. Overall, it provides a known actor for performing user interaction. Therefore, it is the responsibility of the director to spread the information about available IO actors (e.g. at creation time). In addition, it is important that internal communication (between actor, IO actor, and user interface) only includes business object data. It is not the responsibility of the IO actor to create a form for the user. In the contrary, this task is subject to the user interface. 

Internally, in the first step, the process actors request a user interaction from the IO actor. Since users need some form of context, this request further includes information about the instance or state the request originates from. This information is partly hard-coded as well as initially given out by the director on actor startup (see \autoref{Director_concept}). In general, this information also needs to be transformed into some form of human-understandable names (e.g. instance name, model name).
The next step is that some form of interface requests the pending user interactions from the IO actor. The actual user interaction occurs purely in the user interface. Therefore, the IO actor acts as an intermediary between process actors and user interfaces.

\begin{figure}[htb]
	\centering
	\includegraphics[width=0.485\textwidth]{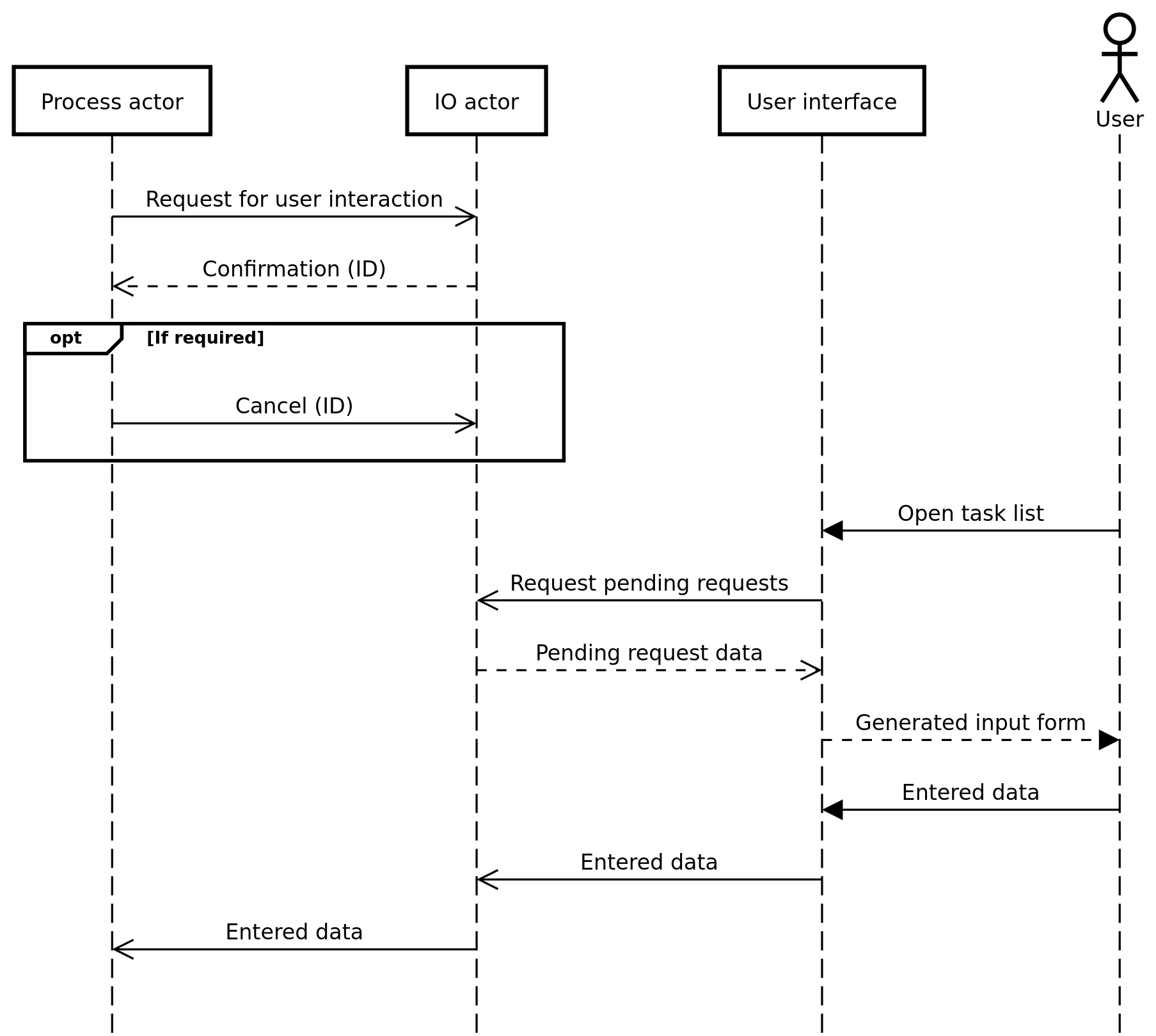}
	\caption{Sequence diagram of IO actor communication and user interaction}
	\label{fig:ioactorconcept}
\end{figure}

After the user has performed the interaction, the interface sends back a request to the IO actor. Afterwards, a lookup on which actor the request belongs to is performed, and the data is sent to the originating actor. As a consequence, the request is served and deleted by the IO actor. In order to identify an interaction request throughout the whole process, IDs are needed. The overall process is depicted as a sequence diagram in \autoref{fig:ioactorconcept}.

Furthermore, it is possible that multiple IO actors exist. The only requirement is that they are all known to the actors. As previously mentioned, it is the task of the director to make IO actors known. A use case for multiple IO actors is to provide a variety of communication techniques. For example, instead of utilising internal communication through the management actor system (and therefore the user interface), web requests or publish and subscribe systems could be used. In such a scenario, some logic has to be in place in order to select the IO actor within the model or at runtime. 

Therefore, the IO actor could also be used to communicate with external systems as it is a requirement for workflow engines (see \autoref{requirements}). For example, it could handle \acrshort{API} calls in a centralised place. As an alternative, this logic could be baked directly into the actor's code, leading to reduced flexibility. For example, it would not be possible to change the address of an \acrshort{API} without modifying the generated code (and also redeploying it).

\subsection{User interface and user roles}
\label{user_interface}

In order to provide a method for users to interact with the system, a user interface is necessary. In addition, as listed in \autoref{requirements}, it has to provide a task list as well as a way for entering and manipulating data (i.e. working with business objects) during the process execution. Furthermore, the user interface does not only act as a method of user interaction but also maps users to its corresponding subject in the model. Therefore, it holds the user information and can also be synchronised with existing directory services such as \acrfull{LDAP} or \acrshort{AD}. This allows for integration into organisational structures. Based on this mapping, work is distributed to the responsible users or groups. The mapping from \acrshort{PASS} subject to user or group could be done manually or automatically based on rules. For example, mapping rules could be based on naming conventions of subjects. 

Furthermore, the user interface also exposes the additional functionality to the users (in respect to the roles described in \autoref{user_interface}). This includes uploading models or creating process instances. Since the user interface is also in charge of integrating with organisational structures, it can restrict functionality based on roles. Therefore, an access control system can be implemented.

Within the interface, at least two main roles are defined for a privilege separation. First is the admin role which is responsible for compiling and loading the models into the actor system and second the actual user which is assigned to a specific subject in the model. This can be done directly or indirectly via group assignment and maps the responsibilities of users to the subject within the model. Depending on the specified actor type, the user is also able to start a new process instance, otherwise, the user is only notified when an interaction is requested by an actor. Therefore, the functions available on the interface vary depending on the user's role.

However, due to the separation of the user interface and server side actor system (shown in \autoref{fig:high_level_concept}), different means of communication between the interface and server side actor system could be chosen. For example, interfacing with web requests would be possible when the server side actor system provides a web service. A separate management actor system would not be necessary in such a scenario.

In addition, the interface component is responsible for storing and parsing the uploaded model further translating it to an actor source code. As a result, it also plays an important role when it comes to traceability and logging. The generated source code is stored twice in the system, once in the interface component (in plain text, for debugging and traceability), and in the actor system itself where it can then be executed in the correct actor system.

\subsection{Logging and traceability}

Logging and traceability are important requirements for workflow engines as described in \autoref{requirements}. The proposed approach includes multiple layers of logging while also providing ways to ensure traceability which are inherent to the code generator approach. First, basic logging functionality can be implemented in the user interface (to log information about user interactions, access logs, etc.). Second is the logging within the actors themselves. Given that a centralised logging actor exists, the code generator can add log statements to the code. Therefore, information about the execution can easily be collected. 

The generated code is stored twice and in plain text. As a result, it is possible to read the code that is executed which can be useful to identify potential problems before they occur. 

\subsection{Security and multi-company considerations}
\label{multienterprise}
As previously stated, multiple distributed actor systems can be interconnected. This does bring benefits when it comes to scalability and redundancy but also opens up the possibility for multi-company deployments of the concept. Conclusively, the execution of process models which include subjects residing in different companies would be possible. The proposed approach assumes that the actors generated from such models are executed on the actor systems running the corresponding companies. Therefore, the placement of actors on actor systems has to be controlled. This scenario is already depicted in \autoref{fig:high_level_concept} with two different companies (Server actor system 1 and 2). 

Furthermore, the internal actors need additional mechanisms in place to support multi-enterprise scenarios. Those mechanisms would have to ensure that no internal information gets shared between companies and companies can be added in a hot-plug scenario. Additionally, user interaction as well as external service calls must be routed correctly.

However, this adds complexity to the internal structure and some security considerations have to be taken. It is assumed that the communication between the actor systems is confidential which can either be assured by the implementation itself (e.g. communicating through encrypted protocols) or by utilising a \acrfull{VPN} \cite{hamedModelingVerificationIPSec2005}.

Furthermore, it has to be ensured that a company cannot introduce and execute malicious code into the other company's actor system. This concept does not include an explicit permission system for a company to select which actors can be started by other companies. A possible solution would be the use of a \acrfull{PKI}, to only allow signed code to be executed.

\section{Workflow engine prototype}
\label{Prototype}
In this section a prototypical implementation of the concept presented in \autoref{section: workflow-concept} is described in detail. Overall the focus is on the basic constructs of \acrshort{PASS} while additional elements are taken into consideration since they are needed for the validation patterns. Extended language elements aim to reduce model complexity but can also be modelled using only the basic constructs \cite{fleischmannWhatSBPM2010}. Therefore, it is beyond the goal of the prototype to support them as well. In \autoref{tbl:summary_supported_elements} a summary of supported elements along with their limitations is given.

\begin{table}[htb]
	\caption{Summary of supported \acrshort{PASS} language elements}
	\label{tbl:summary_supported_elements}
	\centering
	\begin{tabular}{|p{0.09\textwidth}|p{0.35\textwidth}|}
		\hline
		Element & Limitations and restrictions. \\ [0.5ex]
		\hline\hline
		Subjects & Supported, but without multi-subject support. \\
		\hline
		Messages & Supported, priorities and multi-subjects not supported. \\
		\hline
		Business objects & Only basic data types such as integers, strings and dates are supported. \\
		\hline
		Send-states & Supported, but without alternative messages in case the primary one could not be sent and without multi-subject support. \\
		\hline
		Receive-states & Supported, without priorities and multi-subject support. \\
		\hline
		Do-states & Supported, but only with user interaction. \\
		\hline
		Time-transition & Only \texttt{Day\linebreak[0]Time\linebreak[0]Timer\linebreak[0]Transition} is supported. \\
		\hline
	\end{tabular}
\end{table}

For the implementation the programming language Python alongside the frameworks Thespian\footnote{\url{https://pypi.org/project/thespian/}} (for the actor system), Django\footnote{\url{https://www.djangoproject.com/}} (for the web interface) as well as owlready2\cite{lamyOwlreadyOntologyorientedProgramming2017} and lxml\footnote{\url{https://pypi.org/project/lxml/}} (working with ontologies and XML) are used. In general, the structure matches the concept described in \autoref{section: workflow-concept}. For this reason, the inner structure is also depicted in \autoref{fig:detail_concept}.

\subsection{Code generator}
\label{codegenerator}
\begin{figure}[htb]
	\centering
	\includegraphics[width=0.485\textwidth]{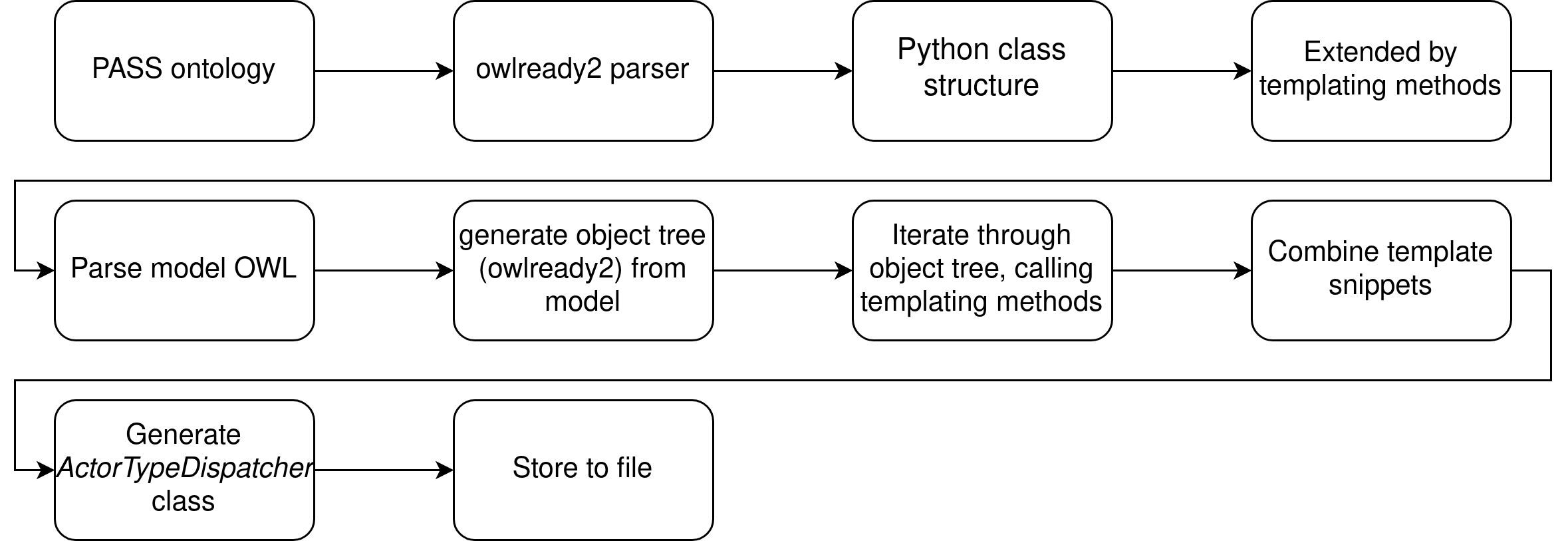}
	\caption{Internal procedure of the code generator implementation}
	\label{fig:code_gen_flow}
\end{figure}

The code generator is responsible for parsing the model uploaded by the user using the web interface and translating it into Python code. According to literature\cite{singerAgentBasedBusinessProcess2016}, approaches utilising the \acrfull{AST} could be used. However, for the sake of simplicity a templating approach is used in the prototype. This approach allows for easy integration with the owlready2 module which basically generates a Python class structure from the \acrshort{PASS} ontology. This generated class structure can then be extended with methods which return template code. Therefore, it is only needed to iterate through the model to call the corresponding method and combine the template snippets into a valid actor source, ready to be used by Thespian. This approach is supported by the general structure of the Thespian actor classes. The steps occurring within the code generator component are depicted in \autoref{fig:code_gen_flow}

In the following sections, the code generated for each supported element in the \acrshort{OWL} is described in detail. This includes the three main states along with time-transitions as well as the initialisation and exiting phases. All \acrshort{PASS} models are created using the Microsoft Visio shapes developed by Matthes Elstermann in \cite{elstermannProposalUsingSemantic2017} and utilise the \acrshort{PASS} language.

\subsubsection{Subjects}

Each subject in the \acrshort{SID} of a supported model is represented by a \texttt{Fully\linebreak[0]Specified\linebreak[0]Subject} in the \acrshort{OWL}. Those subjects are translated into actors which derive from the \texttt{Actor\linebreak[0]Type\linebreak[0]Dispatcher} class, provided by the Thespian framework. As a result, for each subject, a file containing a class is generated. This class contains the actual behaviour of the subject and consists of multiple methods. Those methods represent the state transitions while the logic of the states is contained within those methods. 

A method can only be triggered by an incoming message and its type. Therefore, each method is (by its method name) linked to a data type or class. If a message of a data type is received by the actor, the corresponding method is called. Therefore, the class consists of a set of loosely coupled methods that are named after a data type or class (i.e. \texttt{receiveMsg\_dict} in case of a dictionary). Automatic state changes (e.g. after a send-transition) can be implemented by sending messages to the actor itself.

Furthermore, an annotation is used to ensure proper actor placement on the desired actor system. This ensures that no actors are wrongly placed within the management actor system. In multi-company scenarios, this could be used to ensure the correct placement of actors according to the association with a company. 

\begin{figure} [htb]
	\centering
	\subfloat[General state structure]{%
		\includegraphics[width=0.3\linewidth]{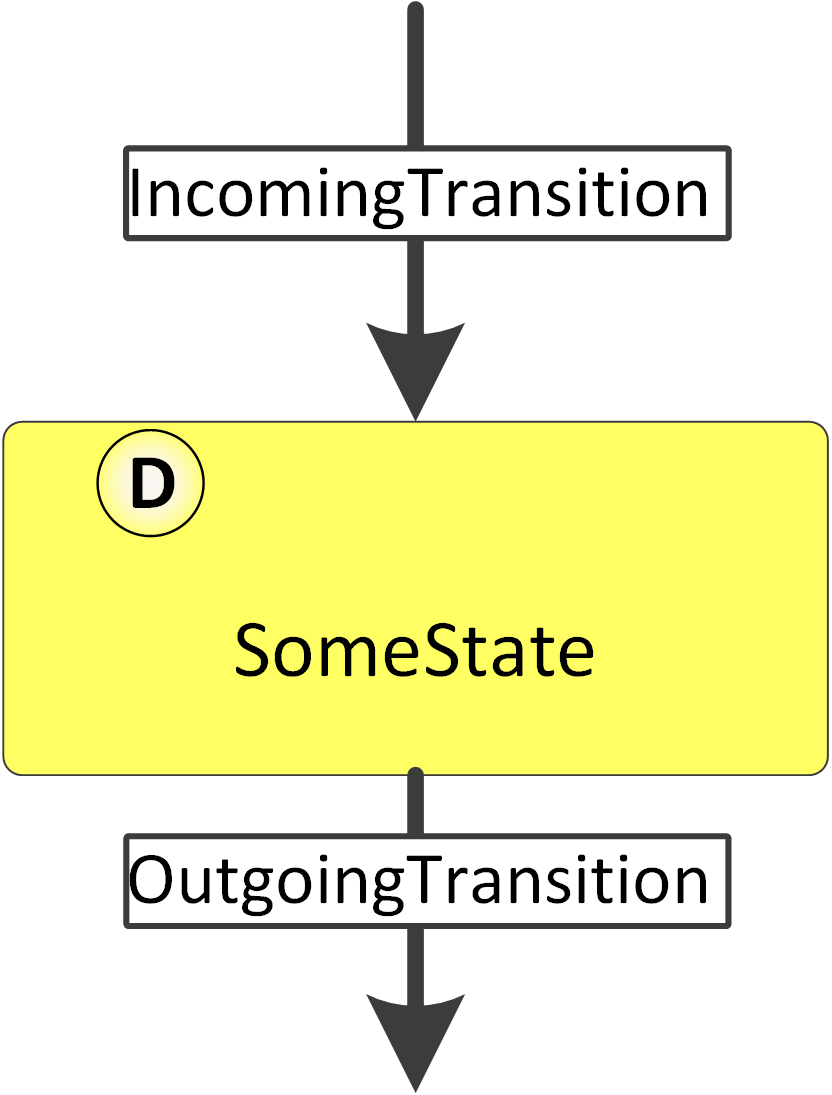}}
	\hfill
	\subfloat[General Python structure generated from a state]{
\begin{lstlisting}[language=Python]
class MyActor:
 func receive_IncomingTransition():
  ExecuteStateLogic(SomeState)
  SendMessage(myself, OutgoingTransition)
\end{lstlisting}  }
\caption{Visualisation and pseudocode of internal actor structure}
\label{fig:except_model}
\end{figure}

To summarise, a mapping of the \acrshort{PASS}-transitions to the \texttt{ActorType\linebreak[0]Dispatcher} class is done by modelling the transitions as messages send to the actor itself. The state logic is included in methods triggered on incoming messages. Therefore, as a general structure, each state within the \acrshort{PASS} model corresponds to a method which is triggered on a defined data type representing the incoming transition and sends a message with the data type corresponding to the outgoing transition to itself. In \autoref{fig:except_model}, this pattern is visualised using pseudocode and a pseudo model.

To enforce a strict order of execution, each state checks if it is currently allowed to be executed. If this is not the case, the method simply exits. However, in case a message from another actor is received, the message is stored in the input-pool for further processing.

\subsubsection{Initialisation and internal message dispatching}
\label{initialization_prototype}

After the initial start, the actor requires information about the environment in order to operate correctly (e.g. address of the director for registration, instance ID). This information is given as a message by the creating actor and is processed in an initialisation routine. This method is triggered when a dictionary data type is received which can originate from the director or another process actor, depending on the logic of the mode.

Furthermore, built-in types (such as dictionaries) are also used to communicate with the director (described in \autoref{Director}) and IO actor (described in \autoref{IOActor}). To differentiate the requests, a type key is contained in the dictionary indicating the message type. This allows reusing the initialisation method for other purposes such as handling the broadcast originating from the director when a new actor registers itself. 

The actual initialisation process consists of storing a set of values received as part of the initialisation message, including the address of the director and IO actor as well as information about the instance such as name and ID. Furthermore, this routine is responsible for registering this actor with the director by its instance ID. After the initialisation routine is done the behaviour continues as described in the \acrshort{SBD}.

\subsubsection{Do-states}
\label{do-state_prototype}

A do-state indicates the execution of an internal action or function. For this prototype, this is assumed to be done by a human for demonstration purposes. Therefore, each do-state results in a user interaction. However, automated tasks can be implemented by injecting arbitrary code instead of a user interaction which could potentially lead to code simplifications. For functions executed by humans, a user interaction is required which the prototype implements in a two step process.

The first method is triggered by the incoming transition data type according to the overall concept (see \autoref{fig:except_model}). This first method is responsible for sending an IO request to the IO actor. Afterwards, a message type is stored locally. The receive method for this type represents the second step within this process.
After the IO actor responds, the initialisation routine is triggered. This is due to the usage of Python dictionaries (see \autoref{initialization_prototype}). Within this routine, a message of the type previously stored is created and sent to the actor itself triggering a state transition.

\begin{figure}[htb]
	\centering
	\includegraphics[width=0.258\textwidth]{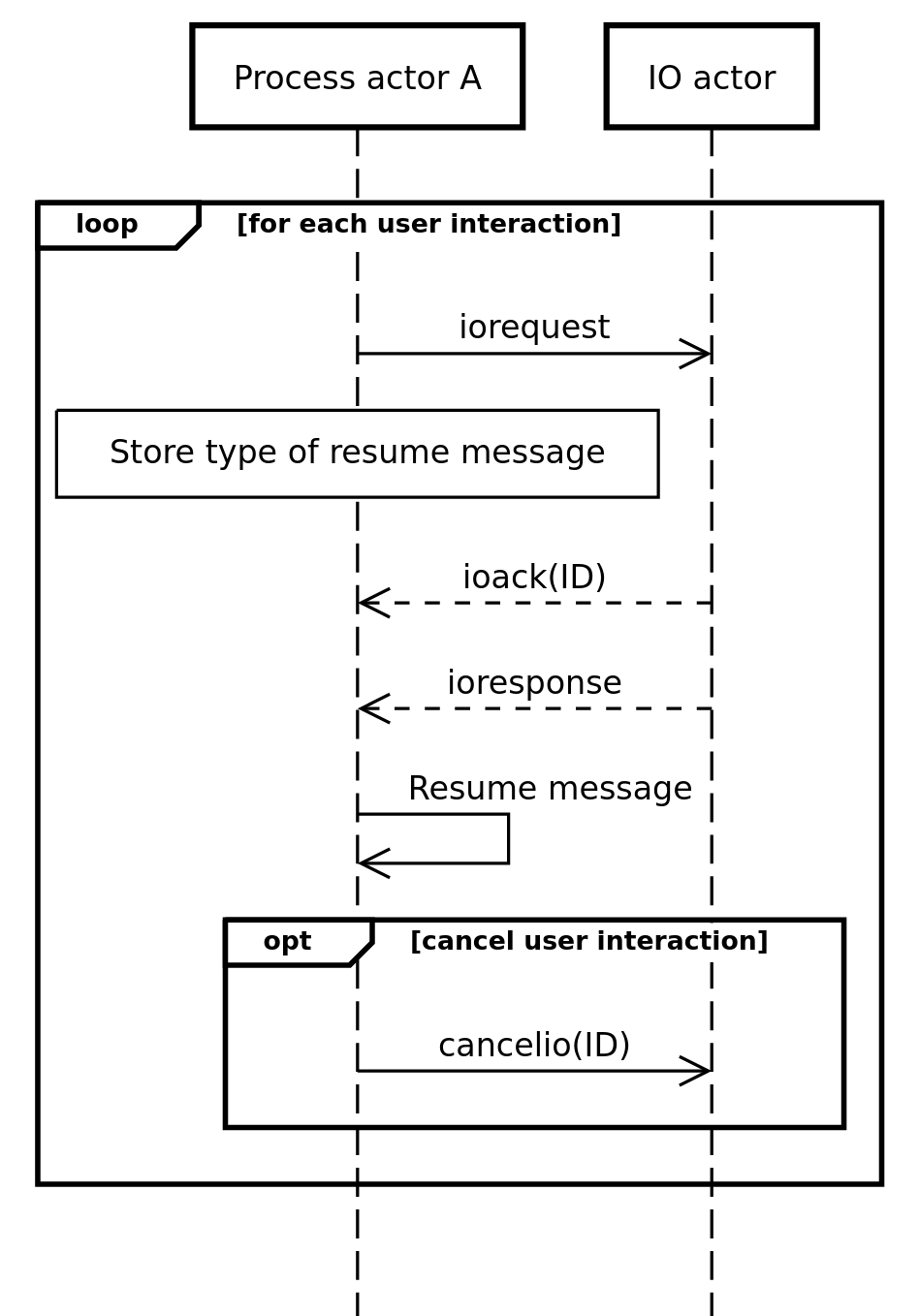}
	\caption{Communication of a process actor with the IO actor}
	\label{fig:io_actor_communication}
\end{figure}

Bound to the receive method of this type (Python data type, not \texttt{dict} at this point), is the second method. Included in this message is a \acrshort{JSON} string representing the user input. This data is stored in local variables according to the data mapping defined in the model and, if multiple transitions are possible, the selected one is triggered. For further clarification, the sequence diagrams \autoref{fig:ioactorconcept} and \autoref{fig:io_actor_communication} are given.

\subsubsection{Receive-states}

The receive-state is a simple construct. It consists of a method which is triggered by a message received from other actors. This matches with the overall structure of the \texttt{ActorTypeDispatcher} class stemming from Thespian. As a consequence, the overall logic is quite simple, since it only includes data mapping and triggering the outgoing transition. However, since it is also possible that the receive-state represents an end-state, it can also trigger an \texttt{ActorExitRequest}.

Furthermore, a receive-state is a method of acquiring the address of another actor. This is used as a fallback in case an "addressbook" request is not received in time, e.g. when the director responds after an initial send-state. As previously mentioned, the receive-state is special when it comes to enforcing execution order.  

In the case of receive-states, the message is stored in a list (representing the input-pool in the \acrshort{PASS} concept) if received outside of the correct state. Afterwards, each receive-state checks if a message of the needed type is already stored in the pool and processes it by sending it to itself. As a consequence, messages are processed in a \acrshort{FIFO} manner, and if no needed message can be found in the pool, the subject goes on to waiting for the required message. In case the message is taken from the pool, the original sender address is not preserved. 

The messages triggering all other states originate from the actor itself. If such a message is received outside an allowed state, the message is simply discarded to ensure a proper order of states.

\subsubsection{Send-states}

\begin{figure}[htb]
	\centering
	\includegraphics[width=0.485\textwidth]{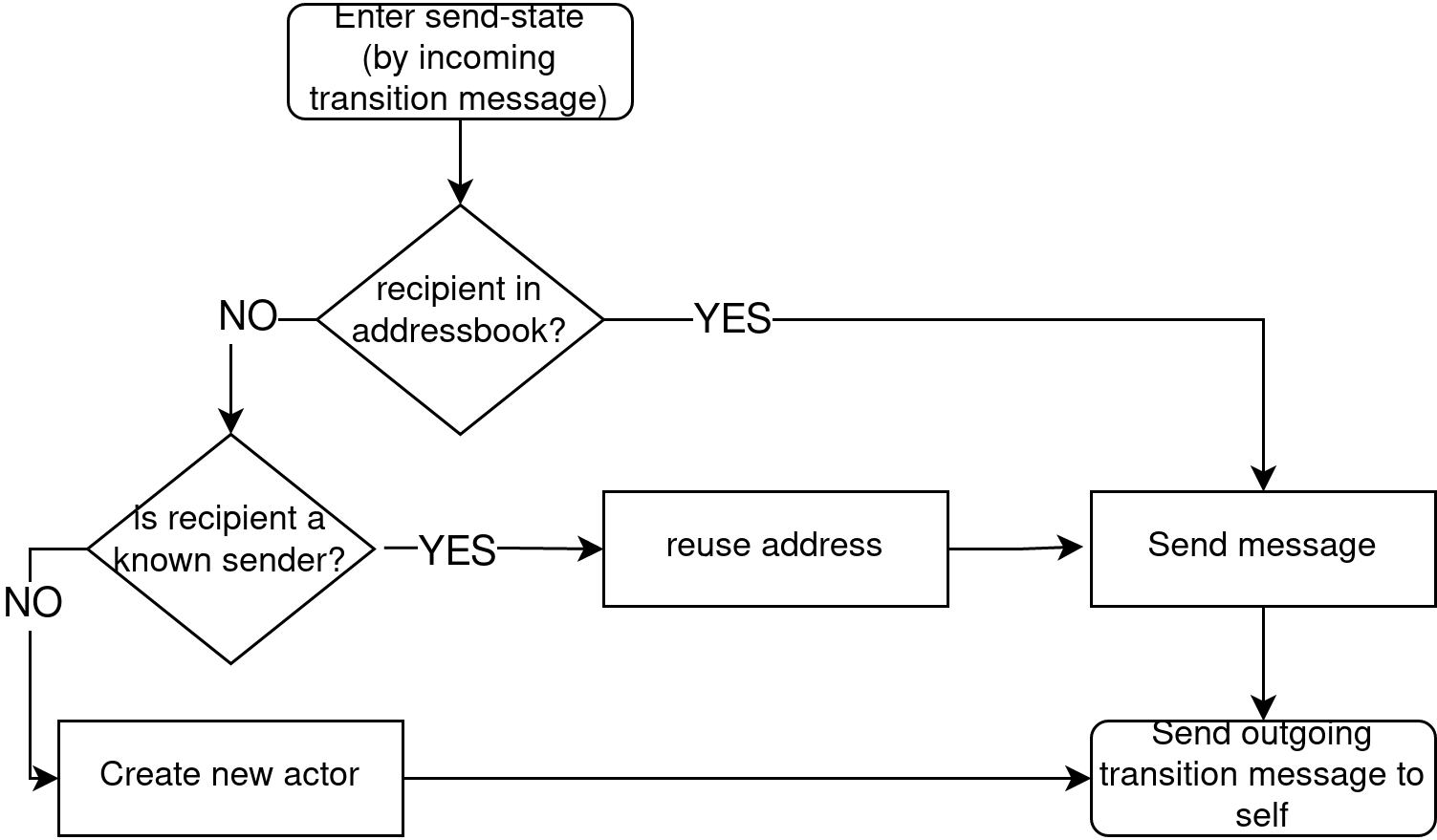}
	\caption{Flowchart of the internal send-state logic}
	\label{fig:flowchart_sendstate}
\end{figure}

A send-state is, as the name implies, responsible for sending messages to another actor. However, it is also the responsibility of a send-state to ensure that the other party exists. Therefore, the send-state generates a message according to the data mapping specified and tries to find the actor address of the recipient. This process is depicted in \autoref{fig:flowchart_sendstate}. At first, it is checked if the recipients address is available locally. Usually, this information is acquired as part of the \texttt{addressbook} request, broadcasted by the director.

If there is no actor address available that was broadcasted before, a local variable is consulted. Actor addresses are stored in this variable when a message is received. This serves as a backup to ensure that a new actor is only created when necessary. Otherwise, a new actor is created and the initialisation message is sent immediately. After the actual message is transmitted, the outgoing transition message is triggered.

\subsubsection{Time-transitions}

Time-transitions can originate from every state and have a non blocking behaviour similar to an IO request. If a state has an outgoing time-transition (the prototype only supports the type \texttt{DayTimeTimerTransition}), the target state type is stored in a global variable and a wake-up message of the Thespian framework is scheduled. This message is received by the \texttt{receiveMsg\_WakeupMessage} method and a message of the type pending is generated.

However, if a time-transition is not triggered, there is no functionality offered by Thespian to cancel a pending wake-up message. In this case, each new state (omitted in all listings) clears the timeout variable. Then, if the wake-up message is received, it is simply ignored, since the pending variable is empty. Similar to an actor exit, all pending IO requests are cancelled if the originating state was a do-state.

\subsubsection{Exiting}

In general, actors are executed until an \texttt{Actor\linebreak[0]Exit\linebreak[0]Request} is received. This message is handled in the same way as other messages, but after the corresponding \texttt{receive\linebreak[0]Msg\_Actor\linebreak[0]Exit\linebreak[0]Request} method, the actor is killed by the actor system. Therefore, this behaviour is used to ensure that all pending IO requests are cancelled and the actor deregisters itself from the director. However, if the actor has created any subsequent actors in its lifetime (child actors) the request is propagated, and they are terminated as well. This behaviour generally does not match the independence of subjects and can be prohibited by setting the recursive flag to false when creating an \texttt{Actor\linebreak[0]Exit\linebreak[0]Request}.

Furthermore, this approach does not fully comply with the \acrshort{PASS} specification where an end-state leads not necessarily to the termination of a subject. Instead, a subject can be reactivated and is only terminated when all other subjects within this process are also in an end-state. This behaviour is not available in the prototype.

\subsubsection{Data handling and classes}

The core concept of the Thespian \texttt{ActorTypeDispatcher} class are methods triggered by the receipt of messages. The exact method depends on the type of the message. Therefore, the types and messages are closely related to the code generator. As previously mentioned, a file is generated for each subject in the model. Besides that, a common file consisting only of data classes is generated as well. This contains classes used for state transitions internally as well as message definitions. Therefore, a message content can simply be modelled by a class and its variables. In case there is no data mapping available for a message, a simple placeholder content is used. Internal messages exhibit the same structure.

In order to generate a form for the user within the interface, additional information such as display name and data type are stored as well. This is needed since Python is a dynamically typed language and types are given in the model. Based on the data type, special input fields such as date-pickers for date and time data are generated.

Data access within states is done by mappings (incoming and outgoing) and data access mappings (within do-states, defining read/write variables). Data mappings simply result in variable assignments in do- and receive-states and a new object instantiation in case of a send-state.

However, a formal method of expressing data mappings is not standardised within the \acrshort{PASS} standard. As a consequence, the \acrshort{XML}-literals used are a result of the internal workings of the Microsoft Visio shapes \cite{elstermannPersonalCommunication2022}.

\subsubsection{Multi-subject support}

Multi-subjects are not directly supported by the code generator. However, if the number of subjects is known while modelling the desired behaviour can be implemented manually. On the execution side, the actor system used does not enforce any limitations on the number of actors running. Therefore, the support of multi-subjects comes down to the code generator component.

In order to implement multi-subjects correctly, certain aspects have to be taken into consideration. This includes questions regarding where actors are started (creating a new one versus reusing an existing one), referring to actors (which instance is the correct recipient? how are references stored?), and how data handling is done. Given that this information is provided in the model as well as in the \acrshort{PASS} definition, an implementation is most likely possible. However, concerning those questions, some aspects may not be fully specified at this point \cite{elstermannPersonalCommunication2022}.

\subsection{Runner}
\label{runner}
The runner is a component within the web interface application and is responsible for administrative interaction between the \acrshort{GUI} and the Thespian actor system. This includes operations such as loading and unloading source code as well as starting and stopping actors. This is done in close coordination and interaction with the director actor described in \autoref{Director}. Internally, the runner starts an actor system itself when called which connects to the server side actor system to establish a way of communication. The runner itself only functions as an interface to the actor system and does not act directly on actors. Instead, it interfaces with the director to perform its operations.

\begin{figure}[htb]
	\centering
	\includegraphics[width=0.485\textwidth]{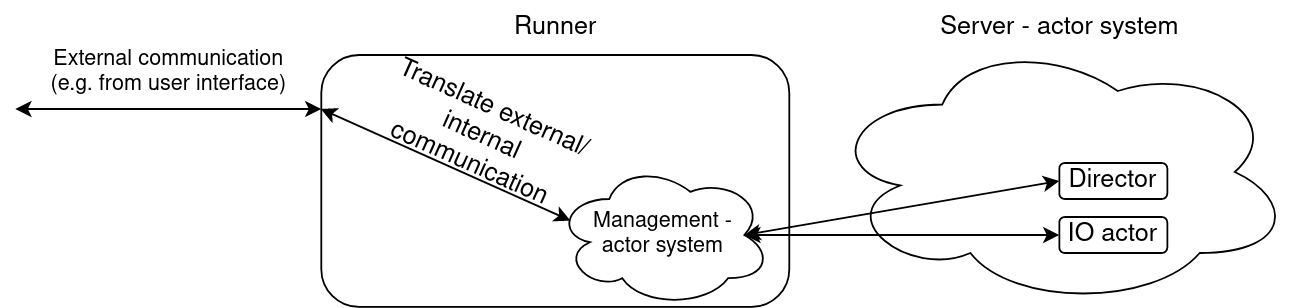}
	\caption{Placement and communication of the runner component}
	\label{fig:structure_runner}
\end{figure}

In addition, the runner is responsible for managing the data transfer between the web interface and the IO actor described in \autoref{IOActor}. All references as well as the actor system object itself is held and managed through a singleton object. This ensures that only one actor system is used for management and the references to the IO actor and director are correct.

A possible security concern arises from the dynamic source loading functionality. It could open up an attack vector where malicious code is introduced into the system. In order to mitigate this problem, Thespian provides the possibility to validate loaded sources via certificates. A reference can be seen in the director component of the framework\footnote{\href{https://thespianpy.com/doc/director.html\#outline-container-orgd6f5567}{https://thespianpy.com/doc/director.html} (Retrieved: 31.07.2022)}. However, it is out of scope for the prototype within this paper to implement such a mechanism.

\subsection{Director}
\label{Director}

The director actor is responsible for holding the information of currently running actors in the system as well as executing the start and stop operations. Therefore, all actors corresponding to process instances can be seen as child of the director actor. With this approach, the director always knows where the actors are located and gets informed by the system if errors are occurring. Its main purpose is to serve as a central point for finding other actors within a process instance (actor discovery, \autoref{Director_concept}). Furthermore, the director actor is highly involved in the initialisation process which is discussed in \autoref{initialization_prototype}.

All actors generated by the code generator described in \autoref{codegenerator} register themselves with the director on startup. This also occurs when they are started by another actor in an instance. In addition, they also send a deregistering message on exit. Since this includes messages across multiple source code versions, only build in data types can be used. In this case, the communication heavily relies on dictionaries (i.e. the Python data type \texttt{dict}). Similar to the generated actors, the director is also based on the \texttt{ActorTypeDispatcher} class.

The director described should not be confused with the integrated director of Thespian which provides similar functionality. For a possible future implementation, this integrated director can be further examined. Furthermore, the director plays an important role in multi-company scenarios. For this use case, further development regarding synchronising multiple instances could be needed. Additional considerations regarding this use-case are described in \autoref{multienterprise}.
\subsection{IO actor}
\label{IOActor}

The IO actor is responsible for holding information about pending interaction requests. Therefore, it is, alongside the director actor, a main point of information for the runner. It consists mainly of two Python dictionaries, the first holding a \acrshort{JSON} string containing the information necessary for the front end, such as data types and input fields. The second one holds the address of the actor requesting the interaction. 

This is linked with a class wide iterator, generating an index which is also sent back to the requesting actor for further referencing this request (\texttt{ioack} message). A common use case for this index used as ID, is the cancellation of interaction requests in case the requesting actor exits, or the state is changed by a time-transition. The overall concept is depicted in \autoref{fig:io_actor_communication}.

\subsection{Web interface}
\label{GUI}

The interface is required to allow for comfortable interaction for users with the system. In addition, it also integrates some of the previously mentioned components such as the runner and code generator. Furthermore, it serves as a repository for compiled models that can be loaded into the actor system on demand. Internally, a database (SQLite for the prototype) is used for storage and a wide variety of databases are supported due to the use of the Django \acrshort{ORM} feature.

\subsection{Logging and monitoring}
\label{logging}

Thespian itself provides a central logging actor. Therefore, each actor is able to write log information which can be written to a file. The exact format as well as log levels can be defined on an actor system level. However, it is the responsibility of the actor itself to decide which information is logged. 

Since the code of the actor is generated, logging statements can be injected into the template code by the code generator. This approach allows for great flexibility considering which, how and when data is logged within a running actor. When it comes to the web interface, the used framework Django also allows integration with standard Python logging tools and stores the models and the generated code for traceability.

\subsection{Additional limitations and possible improvements}

The main goal is to prove the feasibility of the concept developed in \autoref{section: workflow-concept}. Therefore, this prototype exhibits shortcomings and is not intended for real-world use.

A major limitation of the prototype is persistence. The prototype heavily relies on the states and data stored within the actor system and in specific administrative actors (i.e. IO actor, director). Currently, there is no system in place to persist this data in case of a restart. However, mechanisms exist to achieve the desired behaviour as stated in the documentation but are not included in the prototype.

Furthermore, explicit handling of external \acrshort{API} calls is not included. This functionality only needs to be implemented in the code generator since Python is able to make a wide variety of external calls. As a result, the concept is able to handle this requirement without conceptual changes.

The current implementation does not check any message priorities but just operates in a \acrshort{FIFO} manner. Prioritisation could be implemented by checking priorities when the message is taken from the pool or by reordering the pool. This forms basically an additional layer on top of the input queue of Thespian.

Another limitation is the complete support for multiple companies. While the actor systems can be distributed, the director is implemented as a singleton. Therefore, it cannot synchronise with other directors. Furthermore, multiple IO actors are not supported which would be necessary to address the correct user interface.

\section{Possible use case scenarios}
\label{section:results}

Due to the usage of \acrshort{PASS} as an intermediary language a variety of use cases become possible. In the following chapter some of them are tested in a systematic manner. The validation procedure either starts with a \acrshort{BPMN} model which is translated to a \acrshort{PASS} (\acrshort{OWL}) model or a \acrshort{PASS} model generated with the Visio shapes. For \acrshort{BPMN} models exported from Signavio and Camunda are tested. All \acrshort{OWL} files are checked using Protégé \footnote{\url{https://protege.stanford.edu/}} and a reasoner to check if they conform to the standard \acrshort{PASS} ontology. Furthermore, the models used are based on the service interaction patterns described in \cite{barrosServiceInteractionPatterns2005} and \cite{barrosServiceInteractionPatterns2006} and their usage as language benchmark is suggested in \cite{weskeBusinessProcessManagement2007}. The models are further heavily based on the \acrshort{BPMN} interpretations described in the master's thesis of Stefan Raß \cite{rassRequirementsCapabilitiesCloud2013}. In an execution step, the behaviour shown is evaluated based on the descriptions available in literature.

\subsection{Modelling in MS-Visio and execution with coreASM workflow engine}
\label{visioCoreASMexec}

In its original form the coreASM based workflow engine is developed before or in parallel to the \acrshort{OWL} standard. As a result some differences are to be expected. As a further note, the current publicly available version of the \acrshort{ASM} based workflow engine does not support importing \acrshort{OWL} models. However, the engine described in \cite{elstermannMappingExecutionModel2020, wolskiCoreASMBasedReference2019} is able to handle \acrshort{OWL} models and was therefore used for testing. While the tests succeeded and the desired execution behaviour was shown, warnings indicated that enhanced checks performed by the workflow engine were not successful. This could be a result of constraints as described in \cite{PerformanceElstermannWolski} as well as other currently unpublished work.

\subsection{Modelling in MS-Visio and execution with Python workflow engine}

The \acrshort{OWL} exported from MS-Visio was used as a reference when developing and testing the workflow engine prototype. Therefore, the prototype is able to execute the models as expected. However, the MS-Visio \acrshort{OWL} shapes (at the time of development and testing, being also a work in progress) exhibit some specifics which needed to be taken care of in the prototype specifically on how business object definitions are stored.

\subsection{Modelling in BPMN with BPMN tool and execution with Python workflow engine}

As a starting point for modelling \acrshort{BPMN} the tools Signavio and Camunda were used.
Based on the exported \acrshort{XML} the converter was developed.
The resulting \acrshort{OWL} conforms to the standard to such an extend that there where no specific modifications of the workflow engine needed to handle the execution of those models.
In addition, the execution logic was identical to the models generated with the MS-Visio shapes.

\subsection{Modelling in BPMN with BPMN tool and execution with coreASM workflow engine}

As described in \autoref{visioCoreASMexec}, while showing that the workflow engine is able to execute the models under test, some warnings where exhibited which need further testing and research.

An incompatibility occurs within the timer-transition-condition. The coreASM workflow engine requires a \texttt{has\linebreak[0]Duration\linebreak[0]Time\linebreak[0]Out\linebreak[0]Time} attribute with the data type \texttt{xsd:\linebreak[0]duration}. However, the translated models contain this attribute with the data type \texttt{xsd:string}.
Within the PASS ontology both data types are valid for this attribute.
By manually changing the data type the execution of the models with the coreASM workflow engine works.

\subsection{Modelling in BPMN with BPMN tool and continue modelling with MS-Visio}
The MS-Visio shapes are able to import models stored as ontology files by using an Visio plug-in\footnote{\url{https://github.com/MatthesElstermann/ALPS-Visio-Add-In}}. Those models can then be further edited and exported again. However, since the translated models do not contain information about the graphical representation, the graphical representation is generated by the plugin. In some cases this could lead to the need of manual repositioning of elements in order to simplify the further modelling process.

\subsection{Modelling in PASS with MS-Visio and continue modelling with BPMN tool}
The converter is able to translate a \acrshort{PASS} model back to \acrshort{BPMN} which adds another verification layer since it could be compared to the source file. However, information about the graphical representation is lost in the translation process. Therefore it is currently not possibly to re-import the \acrshort{BPMN} file into Signavio or Camunda. The translation of the \acrshort{BPMN} diagram interchange to \acrshort{OWL} and vice versa could be subject of further research.

\section{Conclusion}
\label{section:conclusion}
In this paper, a feasibility study is described which aims to determine to which extent workflow engines are able to execute control flow logic specified in various modelling paradigms. This is an important aspect if parts of a process are running in different organisations. If process models can be specified by different modelling languages and corresponding tools, organisations can use their preferred methods and tools. Especially if complex cross-organisational processes have to be described, each organisation can apply its existing toolset if a common storage format is used. Based on that storage format different execution platforms can be applied in the various involved organisations.\\
In this study as paradigms under test the industry standard modelling languages \acrshort{BPMN} and \acrshort{PASS} have been chosen. The first one due to its widespread adoption and the later for its formally described execution semantics.
\\
To fill this gap between the languages a concept was developed to allow both modelling paradigms to be interchangeably used when it comes to execution of the control flow. 
This was done by defining \acrshort{PASS} as an intermediary layer due to its more formal specification.
Therefore, in the first step, a translator from \acrshort{BPMN} in its XML form to \acrshort{PASS} in its \acrshort{OWL} form was designed and implemented.
Due to structural differences between the two paradigms, only a subset of elements have been specified to be translatable. As a result of this translation, the well-known \acrshort{BPMN} syntax is enhanced by the formally described \acrshort{PASS} execution semantics. Furthermore, the currently only available tool for modelling \acrshort{PASS} was verified to be able to import the translated \acrshort{PASS} files correctly. Both, translated files as well as directly modelled \acrshort{PASS} files were then executed using two different \acrshort{PASS} based workflow engines. The first one is a reference implementation of the formal \acrshort{PASS} \acrshort{ASM} semantics which already exists. The second one is a workflow engine also designed and developed in this paper based around translating the \acrshort{OWL} model to executable Python code.\\
The results show that the approach of \acrshort{PASS} as in intermediary language between workflow engines could be feasible to streamline interchangeability of execution logic between workflow engines.
\\
In the execution of business processes people, software, physical devices and combinations of these implementation techniques are involved. In our feasibility study, only software and people as implementation technology are considered. The integration of physical devices in business process is the subject of further investigations. A general approach to resource management and authorisation is described in \cite{Resource-Management-GBAC} \cite{Role-and-Right-Management-GBAC}. In further research, it has to be investigated how these concepts support method- and tool-independent deployment of business processes across various organisations.



\bibliographystyle{IEEEtran}
\bibliography{IEEEabrv,preprint2.bib}

\begin{thebibliography}{10}
\providecommand{\url}[1]{#1}
\csname url@samestyle\endcsname
\providecommand{\newblock}{\relax}
\providecommand{\bibinfo}[2]{#2}
\providecommand{\BIBentrySTDinterwordspacing}{\spaceskip=0pt\relax}
\providecommand{\BIBentryALTinterwordstretchfactor}{4}
\providecommand{\BIBentryALTinterwordspacing}{\spaceskip=\fontdimen2\font plus
\BIBentryALTinterwordstretchfactor\fontdimen3\font minus
  \fontdimen4\font\relax}
\providecommand{\BIBforeignlanguage}[2]{{%
\expandafter\ifx\csname l@#1\endcsname\relax
\typeout{** WARNING: IEEEtran.bst: No hyphenation pattern has been}%
\typeout{** loaded for the language `#1'. Using the pattern for}%
\typeout{** the default language instead.}%
\else
\language=\csname l@#1\endcsname
\fi
#2}}
\providecommand{\BIBdecl}{\relax}
\BIBdecl

\bibitem{Formal-Methods-Software-Engineering-2023}
R.~Markus, C.~Antonio, S.~Bernd-Holger, S.~Gerardo, and S.~Siraj, Ahmed,
  \emph{Formal Methods for Software Engineering: Languages, Methods,
  Application Domains (Texts in Theoretical Computer Science. An EATCS
  Series)}.\hskip 1em plus 0.5em minus 0.4em\relax {Springer Berlin
  Heidelberg}, 2023.

\bibitem{BPMN-OMG-2012}
{OMG Working Group}, ``{{BPMN}} 2 {{Business Process Model and Notation}}
  ({{Second Edition}}),'' {OMG Recommendation}, Tech. Rep., 2014.

\bibitem{w3cowlworkinggroupOWLWebOntology2012}
{W3C OWL Working Group}, ``{{OWL}} 2 {{Web Ontology Language Document
  Overview}} ({{Second Edition}}),'' {W3C Recommendation}, Tech. Rep., 2012.

\bibitem{BoergerABstractStateMachine2003}
E.~Börger and S.~Robert, \emph{Abstract State Machines: A Method for
  High-Level System Design and Analysis}.\hskip 1em plus 0.5em minus
  0.4em\relax {Springer Berlin Heidelberg}, 2003.

\bibitem{singer2019ontological}
R.~Singer, ``An ontological analysis of business process modeling and
  execution,'' 2019.

\bibitem{BraunauerTimTobias2022BzP:}
T.~T. Braunauer, ``\BIBforeignlanguage{ger}{{BPMN zu PASS : Übersetzung von
  Geschäftsprozessmodellen}},'' Master's thesis, FH JOANNEUM - University of
  Applied Sciences, 2022.

\bibitem{ZeislerGerhard2022Acgf}
G.~Zeisler, ``\BIBforeignlanguage{eng}{Automatic code generation from business
  process models},'' Master's thesis, FH JOANNEUM - University of Applied
  Sciences, 2022.

\bibitem{KurzExchangeStandard2016}
\BIBentryALTinterwordspacing
M.~Kurz, ``Bpmn model interchange: The quest of interoperability,'' in
  \emph{Proceedings of the 8th {{International Conference}} on
  {{Subject-Oriented Business Process Management}} - {{S-BPM One}}}.\hskip 1em
  plus 0.5em minus 0.4em\relax {ACM Press}, 2016. [Online]. Available:
  \url{https://dl.acm.org/doi/abs/10.1145/2882879.2882886}
\BIBentrySTDinterwordspacing

\bibitem{BPMN-Model-Interchange}
\BIBentryALTinterwordspacing
B.~M. I.~W. Group, ``2022 model interchange capability demo,'' 2023. [Online].
  Available: \url{start [BPMN Model Interchange Working Group] (omgwiki.org)}
\BIBentrySTDinterwordspacing

\bibitem{KossakIGKNZKFS14}
\BIBentryALTinterwordspacing
F.~Kossak, C.~Illibauer, V.~Geist, J.~Kubovy, C.~Natschl{\"{a}}ger,
  T.~Ziebermayr, T.~Kopetzky, B.~Freudenthaler, and K.~Schewe, \emph{A Rigorous
  Semantics for {BPMN} 2.0 Process Diagrams}.\hskip 1em plus 0.5em minus
  0.4em\relax Springer, 2014. [Online]. Available:
  \url{https://doi.org/10.1007/978-3-319-09931-6}
\BIBentrySTDinterwordspacing

\bibitem{WikipediaBPMNengines}
\BIBentryALTinterwordspacing
Wikipedia, ``List of bpmn 2.0 engines,'' 2023. [Online]. Available:
  \url{https://en.wikipedia.org/wiki/ List_of_BPMN_2.0_engines, last access
  June 2023}
\BIBentrySTDinterwordspacing

\bibitem{fleischmannPrimerSubjectOrientedBusiness2012}
\BIBentryALTinterwordspacing
A.~Fleischmann, W.~Schmidt, and C.~Stary, ``A {{Primer}} to {{Subject-Oriented
  Business Process Modeling}},'' in \emph{S-{{BPM ONE}} – {{Scientific
  Research}}}, ser. Lecture {{Notes}} in {{Business Information Processing}},
  C.~Stary, Ed.\hskip 1em plus 0.5em minus 0.4em\relax {Springer Berlin
  Heidelberg}, 2012, vol. 104, pp. 218--240. [Online]. Available:
  \url{http://link.springer.com/10.1007/978-3-642-29133-3_14}
\BIBentrySTDinterwordspacing

\bibitem{fleischmannSubjectOrientedBusinessProcess2012a}
\BIBentryALTinterwordspacing
A.~Fleischmann, W.~Schmidt, C.~Stary, S.~Obermeier, and E.~Börger,
  \emph{Subject-{{Oriented Business Process Management}}}.\hskip 1em plus 0.5em
  minus 0.4em\relax {Springer Berlin Heidelberg}, 2012. [Online]. Available:
  \url{http://link.springer.com/10.1007/978-3-642-32392-8}
\BIBentrySTDinterwordspacing

\bibitem{coreASM}
R.~Farahbod, V.~Gervasi, and U.~Glässer, ``An extensible asm execution
  engine,'' in \emph{Proc. 12th International Workshop on Abstract State
  Machines}, 2005, p. 153–165.

\bibitem{Logic-Computation}
M.~Elstermann, A.~Wolski, a.~Fleischmann, C.~Stary, and S.~Borgert, ``The
  combined use of the web ontology language (owl) and abstract state machines
  (asm) for the definition of a description language for business processes,''
  in \emph{Logic, Computation and Rigorous Methods}.\hskip 1em plus 0.5em minus
  0.4em\relax Springer, 2021, pp. 283--300.

\bibitem{article:S-BPM-vs-BPMN-empirical-evaluation}
H.~Moattar, W.~Bandara, M.~Rosemann, and U.~Kannengiesser, ``Are standards
  always best for process modelling? an exploration of bpmn vs s-bpm,''
  \emph{submitted for publication}, 2021.

\bibitem{book:Luhmann}
N.~Luhmann, \emph{Social Systems}.\hskip 1em plus 0.5em minus 0.4em\relax
  Suhrkamp Verlag, 1984.

\bibitem{book:CCS}
R.~Milner, \emph{Communication and Concurrency}.\hskip 1em plus 0.5em minus
  0.4em\relax Prentice Hall, 1989.

\bibitem{book:CSP}
A.~Hoare, \emph{Communicating Sequential Processes}.\hskip 1em plus 0.5em minus
  0.4em\relax Prentice Hall, 1985.

\bibitem{book:Habermas}
J.~Habermas, \emph{Theory of Communicative Action Volume 1, Volume 2}.\hskip
  1em plus 0.5em minus 0.4em\relax Suhrkamp Paperback Science, 1981.

\bibitem{singerAgentBasedBusinessProcess2016}
\BIBentryALTinterwordspacing
R.~Singer, ``Agent-{{Based Business Process Modeling}} and {{Execution}}:
  {{Steps Towards}} a {{Compiler-Virtual Machine Architecture}},'' in
  \emph{Proceedings of the 8th {{International Conference}} on
  {{Subject-oriented Business Process Management}}}.\hskip 1em plus 0.5em minus
  0.4em\relax {ACM}, 2016, pp. 1--10. [Online]. Available:
  \url{https://dl.acm.org/doi/10.1145/2882879.2882880}
\BIBentrySTDinterwordspacing

\bibitem{hamedModelingVerificationIPSec2005}
\BIBentryALTinterwordspacing
H.~Hamed, E.~Al-Shaer, and W.~Marrero, ``Modeling and {{Verification}} of
  {{IPSec}} and {{VPN Security Policies}},'' in \emph{{{13TH IEEE International
  Conference}} on {{Network Protocols}} ({{ICNP}}'05)}.\hskip 1em plus 0.5em
  minus 0.4em\relax {IEEE}, 2005, pp. 259--278. [Online]. Available:
  \url{http://ieeexplore.ieee.org/document/1544626/}
\BIBentrySTDinterwordspacing

\bibitem{fleischmannWhatSBPM2010}
\BIBentryALTinterwordspacing
A.~Fleischmann, ``What {{Is S-BPM}}?'' in \emph{S-{{BPM ONE}} – {{Setting}}
  the {{Stage}} for {{Subject-Oriented Business Process Management}}}, ser.
  Communications in {{Computer}} and {{Information Science}}, H.~Buchwald,
  A.~Fleischmann, D.~Seese, and C.~Stary, Eds.\hskip 1em plus 0.5em minus
  0.4em\relax {Springer Berlin Heidelberg}, 2010, vol.~85, pp. 85--106.
  [Online]. Available:
  \url{http://link.springer.com/10.1007/978-3-642-15915-2_7}
\BIBentrySTDinterwordspacing

\bibitem{lamyOwlreadyOntologyorientedProgramming2017}
\BIBentryALTinterwordspacing
J.-B. Lamy, ``Owlready: {{Ontology-oriented}} programming in {{Python}} with
  automatic classification and high level constructs for biomedical
  ontologies,'' \emph{Artificial Intelligence in Medicine}, vol.~80, pp.
  11--28, 2017. [Online]. Available:
  \url{https://linkinghub.elsevier.com/retrieve/pii/ S0933365717300271}
\BIBentrySTDinterwordspacing

\bibitem{elstermannProposalUsingSemantic2017}
\BIBentryALTinterwordspacing
M.~Elstermann, ``Proposal for {{Using Semantic Technologies}} as a {{Means}} to
  {{Store}} and {{Exchange Subject-Oriented Process Models}},'' in
  \emph{Proceedings of the 9th {{Conference}} on {{Subject-oriented Business
  Process Management}}}.\hskip 1em plus 0.5em minus 0.4em\relax {ACM}, 2017,
  pp. 1--9. [Online]. Available:
  \url{https://dl.acm.org/doi/10.1145/3040565.3040573}
\BIBentrySTDinterwordspacing

\bibitem{elstermannPersonalCommunication2022}
------, Personal Communication, 2022.

\bibitem{barrosServiceInteractionPatterns2005}
A.~Barros, M.~Dumas, and A.~H.~M. ter Hofstede, ``Service interaction
  patterns,'' in \emph{Business Process Management}, W.~M.~P. van~der Aalst,
  B.~Benatallah, F.~Casati, and F.~Curbera, Eds.\hskip 1em plus 0.5em minus
  0.4em\relax Berlin, Heidelberg: Springer Berlin Heidelberg, 2005, pp.
  302--318.

\bibitem{barrosServiceInteractionPatterns2006}
A.~Barros, M.~Dumas, and A.~Ter, ``Service interaction patterns: Towards a
  reference framework for service-based business process interconnection,''
  p.~23, 01 2006.

\bibitem{weskeBusinessProcessManagement2007}
M.~Weske, \emph{Business {{Process Management}}: {{Concepts}}, {{Languages}},
  {{Architectures}}}, ser. {{SpringerLink}}: {{Springer}} e-{{Books}}.\hskip
  1em plus 0.5em minus 0.4em\relax {Springer-Verlag Berlin Heidelberg}, 2007.

\bibitem{rassRequirementsCapabilitiesCloud2013}
S.~Raß, ``Requirements for and {{Capabilities}} of {{Cloud Based BPMS}},''
  Master Thesis, FH JOANNEUM - University of Applied Sciences, 2013.

\bibitem{elstermannMappingExecutionModel2020}
\BIBentryALTinterwordspacing
M.~Elstermann and A.~Wolski, ``Mapping {{Execution}} and {{Model Semantics}}
  for {{Subject-Oriented Process Models}},'' in \emph{Subject-{{Oriented
  Business Process Management}}. {{The Digital Workplace}} – {{Nucleus}} of
  {{Transformation}}}, ser. Communications in {{Computer}} and {{Information
  Science}}, M.~Freitag, A.~Kinra, H.~Kotzab, H.-J. Kreowski, and K.-D. Thoben,
  Eds.\hskip 1em plus 0.5em minus 0.4em\relax {Springer International
  Publishing}, 2020, vol. 1278, pp. 46--59. [Online]. Available:
  \url{https://link.springer.com/10.1007/978-3-030-64351-5_4}
\BIBentrySTDinterwordspacing

\bibitem{wolskiCoreASMBasedReference2019}
\BIBentryALTinterwordspacing
A.~Wolski, S.~Borgert, and L.~Heuser, ``A {{CoreASM}} based reference
  implementation for subject-oriented business process management execution
  semantics,'' in \emph{Proceedings of the 11th {{International Conference}} on
  {{Subject-Oriented Business Process Management}} - {{S-BPM ONE}} '19}.\hskip
  1em plus 0.5em minus 0.4em\relax {ACM Press}, 2019, pp. 1--15. [Online].
  Available: \url{http://dl.acm.org/citation.cfm?doid=3329007.3329018}
\BIBentrySTDinterwordspacing

\bibitem{PerformanceElstermannWolski}
M.~Elstermann and A.~Wolski, ``\BIBforeignlanguage{english}{Performance
  investigation and proposal for updates on the exchange standard for pass},''
  in \emph{\BIBforeignlanguage{english}{Subject-Oriented Business Process
  Management. The Digital Workplace {\textendash } Nucleus of Transformation:
  12th International Conference, S-BPM ONE 2020, Bremen, Germany, December 2-3,
  2020, Proceedings. Ed.: M. Freitag}}, ser. Communications in Computer and
  Information Science (CCIS), vol. 1278.\hskip 1em plus 0.5em minus 0.4em\relax
  {Springer}, 2020, pp. 33--45.

\bibitem{Resource-Management-GBAC}
A.~Lawall, T.~Schaller, and D.~Reichelt, ``Resource management and
  authorization for cloud services,'' in \emph{Proceedings of the 7th
  {{International Conference}} on {{Subject-Oriented Business Process
  Management}} - {{S-BPM One}}}.\hskip 1em plus 0.5em minus 0.4em\relax {ACM
  Press}, 2015.

\bibitem{Role-and-Right-Management-GBAC}
------, ``Role and rights management (1 ed.),'' in \emph{S-BPM in the Wild:
  Practical Value Creation}.\hskip 1em plus 0.5em minus 0.4em\relax Springer
  Verlag, 2015.

\end{thebibliography}

\end{document}